\documentclass[aps,prc,preprint,showpacs,showkeys,superscriptaddress,amsmath,floatfix]{revtex4}

\usepackage{graphicx,natbib}
\usepackage{dcolumn}
\usepackage{bm}

\def\pmb#1{\setbox0=\hbox{#1}%
  \kern-.025em\copy0\kern-\wd0 
  \kern.05em\copy0\kern-\wd0
  \kern-.025em\raise.0433em\box0 }

\newcommand{\Figuretable}[1]{%
  \begin{center} --------- {\bf #1} --------- \\ \end{center}} 
\def\lambdabar{\protect\@lambdabar}
\def\@lambdabar{%
\relax
\bgroup
\def\@tempa{\hbox{\raise.73\ht0
\hbox to0pt{\kern.25\wd0\vrule width.5\wd0
height.1pt depth.1pt\hss}\box0}}%
\mathchoice{\setbox0\hbox{$\displaystyle\lambda$}\@tempa}%
{\setbox0\hbox{$\textstyle\lambda$}\@tempa}%
{\setbox0\hbox{$\scriptstyle\lambda$}\@tempa}%
{\setbox0\hbox{$\scriptscriptstyle\lambda$}\@tempa}%
\egroup
}

\begin{document}


\title{
Coulomb-assisted $\bm\Sigma^-$-nucleus bound states \\
in the ($K^-$,~$\pi^+$) reaction
}


\author{Toru Harada}
\affiliation{%
Research Center for Physics and Mathematics, 
Osaka Electro-Communication University,\\
 Neyagawa, Osaka, 572-8530, Japan
}

\author{Yoshiharu Hirabayashi}%
\affiliation{%
Information Initiative Center, 
Hokkaido University, Sapporo, 060-0811, Japan
}


\date{\today}

\begin{abstract}
We study a production of Coulomb-assisted $\Sigma^-$-nucleus bound 
states by nuclear ($K^-$,~$\pi^+$) reactions
within a distorted-wave impulse approximation, 
so as to examine several types of the $\Sigma$-nucleus potentials
that are consistent with the available $\Sigma^-$ atomic X-ray data 
and nuclear ($\pi^-$,~$K^+$) data. 
We theoretically demonstrate the inclusive ($K^-$,~$\pi^+$) spectra 
of the $\Sigma^-$ unstable bound states 
on $^{28}$Si, $^{58}$Ni, and $^{208}$Pb targets
at incident $K^-$ lab momenta $p_{K}= 400\text{--}800$ MeV/c. 
The results show that the near-recoilless ($K^-$,~$\pi^+$) reaction 
on the $^{58}$Ni target gives a clear candidate to confirm properties 
of the $\Sigma$-nucleus potentials having a repulsion inside the nuclear 
surface and an attraction outside the nucleus with a sizable absorption, 
whereas details of the repulsion of the potential at the nuclear center 
cannot be determined by the inclusive spectra.
This is a promising attempt to extract properties of the $\Sigma$-nucleus 
potential in the nucleus at forthcoming J-PARC experiments, 
as a full complement to the analyses of the $\Sigma^-$ atomic 
and ($\pi^-$,~$K^+$) data. 

\end{abstract}



\maketitle

\section{Introduction\label{sect1}}

It has been discussed that a study of a negatively charged $\Sigma^-$ 
hyperon in nuclei would provide valuable information concerning 
the maximal mass of neutron stars \cite{Balberg1997},
in which a baryon fraction is found to depend on properties of 
hypernuclear potentials in neutron stars. 
The $\Sigma^-$ hyperon in nuclei undergoes a fast 
decay via strong $\Sigma N \to \Lambda N$ conversion processes
due to the mass difference of $m_\Sigma - m_\Lambda \simeq 80$ MeV.
Gal and Dover \cite{Gal1980} estimated a broad 
$\Sigma^-$ width of $\varGamma_\Sigma \simeq 23$ MeV in nuclear matter
by effectively describing the conversion processes 
as the imaginary part of a $\Sigma$-nucleus (optical) potential. 
One of the most important subjects in this research field 
is to clarify properties of the real and imaginary parts of 
the $\Sigma$-nucleus potential. 

The latest analyses of strong-interaction shifts 
and widths in $\Sigma^-$ atoms, 
which are obtained from the $\Sigma^-$ atomic X-ray data, 
have suggested that the $\Sigma$-nucleus potential has a repulsion 
inside the nucleus and a shallow attraction outside the nuclear surface
\cite{Friedman2007}. 
However, it should be noticed that the $\Sigma^-$ atomic energies 
and widths are not so sensitive to a radial distribution of 
the $\Sigma$-nucleus potential inside the nucleus.

Noumi and his collaborators \cite{Noumi2002,Saha2004} 
have performed measurements of $\Sigma$-hypernuclear production by 
inclusive ($\pi^-$,~$K^+$) reactions on C, Si, Ni, In and Bi 
targets at $p_{\pi}=$ 1.20 GeV/c in KEK-E438 experiments. 
Several theoretical analyses 
have also suggested that a repulsive component in the 
$\Sigma$-nucleus potentials is needed to reproduce 
the observed spectra of ($\pi^-$,~$K^+$) reactions 
on nuclear targets \cite{Saha2004,kohno2004,Harada2005}. 
This repulsion originates from the $\Sigma N$ $T=3/2$, $^3S_1$ 
channel, of which state corresponds to a quark Pauli-forbidden 
state in the baryon-baryon system \cite{Fujiwara2007,Rijken2008},
and it is a candidate for the appearance of quark degrees 
of freedom in nuclear physics. 

In a previous paper \cite{Harada2005}, we have succeeded 
to explain the $^{28}$Si($\pi^-$,~$K^+$) data 
as well as the $\Sigma^-$ atomic X-ray data simultaneously, 
by using the $\Sigma$-nucleus potentials that have a repulsion 
inside the nuclear surface and an attraction outside 
the nucleus with a sizable absorption \cite{Batty1994}. 
Nevertheless, the radial distribution of the potential
inside the nucleus and its strength at the center are hardly determined
by fits to the $\Sigma^-$ QF spectrum \cite{Harada2005,Harada2006a}.
Since the $\Sigma^-$ continuum states in the QF region are favored 
in the ($\pi^-$,~$K^+$) reaction 
due to its large momentum transfer of ${\sim}400$ MeV/c, 
quantitative ambiguity of the potential 
cannot be resolved in the analysis 
of the complicated continuum states over a 
wide excitation-energy range \cite{Harada2005}.

One expects that there are the $\Sigma^-$-nucleus 
bound states assisted by the Coulomb attraction even if 
the $\Sigma$-nucleus potential is repulsive.
The wave functions of the $\Sigma^-$ states 
are sizably moderated by strong interactions
because a rms radius for a $\Sigma^-$ atomic $1s$ state in 
medium-to-heavy nuclei is comparable to its nuclear size, 
e.g., $\langle r^2 \rangle^{1/2}_{1s}=$ 4.2 fm 
for a $^{57}$Co core-nucleus where $R=1.1A^{1/3}=$ 4.23 fm.
Yamazaki et al. \cite{Yamazaki1988} called 
these states ``Coulomb-assisted hybrid bound states'', 
rather than $\Sigma^-$ atomic states.
Can we clearly observe such a $\Sigma^-$ bound state? 

In this paper, we theoretically demonstrate a production 
of the Coulomb-assisted $\Sigma^-$-nucleus bound states 
by nuclear ($K^-$,~$\pi^+$) reactions for 
forthcoming J-PARC experiments. 
It is well known that the nuclear ($K^-$,~$\pi^+$) reaction 
provides the recoilless condition 
for a $\Sigma^-$ production (see Fig.~\ref{fig:1}),
which leads to the optimum population of a $\Delta L=0$ transition 
on nuclear targets \cite{Dalitz1978}. 
For medium-to-heavy nuclei, however, 
it seems that individual $\Sigma^-$ levels with broad widths 
are unseparated because the level densities are high 
in the $\Sigma^-$ bound region.
Thus an appropriate momentum transfer near recoilless conditions
is required to selectively populate a non-substitutional bound state
in $\Sigma^-$-nucleus systems.
We perform a calculation of the inclusive spectra on 
$^{28}\text{Si}$, $^{58}\text{Ni}$ and $^{208}\text{Pb}$ 
targets within a distorted-wave impulse approximation (DWIA)
in order to examine properties of the $\Sigma$-nucleus 
potentials. 
We attempt to extract quantitative information on the repulsive 
component of the $\Sigma$-nucleus potentials
from the calculated ($K^-$,~$\pi^+$) spectra.
This is a natural extension of examinations of 
the $\Sigma$-nucleus potentials by $\Sigma^-$ production 
reactions on nuclear targets \cite{Harada1995b,Harada2005,Harada2006a}.

The outline of this paper is as follows:
In Sect.~\ref{sect:theory}, we briefly mention 
a framework for the nuclear ($K^-$,~$\pi^+$) reactions in a DWIA. 
In Sect.~\ref{sect:pot}, we show properties of several 
$\Sigma$-nucleus potentials that are consistent with 
the $\Sigma^-$ atomic X-ray data 
\cite{Friedman2007}
and nuclear ($\pi^-$,~$K^+$) data \cite{Noumi2002,Saha2004}.
In Sect.~\ref{sect:bound}, we calculate the binding energies and widths 
of the Coulomb-assisted $\Sigma^-$-nucleus bound states for 
$\Sigma^-$--$^{27}\text{Al}$, $\Sigma^-$--$^{57}\text{Co}$
and $\Sigma^-$--$^{207}\text{Tl}$ systems. 
In Sect.~\ref{sect:results}, 
we show numerical results of the inclusive spectra in the 
$\Sigma^-$ bound region by the ($K^-$,~$\pi^+$) reactions 
on $^{28}$Si, $^{58}$Ni, and $^{208}$Pb targets 
at $p_K= 400\text{--}800$ MeV/c and $\theta_{\rm lab}=5^\circ$.
We discuss the dependence of a peak structure in the spectra 
on various types of the $\Sigma$-nucleus potentials 
in order to discriminate among these potentials.
Summary and conclusion are given in Sect.~\ref{sect:summary}.

\section{Theory\label{sect:theory}}

\subsection{Distorted-wave impulse approximation (DWIA)}

Hypernuclear production cross sections have been usually calculated 
with the framework of a DWIA 
\cite{Hufner1974,Bouyssy1977,Auerbach1983,Dover1980,
Dover1983,Motoba1988}. 
The double-differential cross section for the ($K^-$,~$\pi^+$) 
reaction at a forward-direction angle
$\theta_{\rm lab}$ in a lab frame is written \cite{Morimatsu1988,Dover1989} by
\begin{equation}
{{\displaystyle d^2 \sigma} \over {\displaystyle d E_{\pi}d \Omega_{\pi}}}
= \beta {1 \over [J_A]}\sum_{M_A}
\sum_{B, M_B} |\langle {\varPsi}_B| \, \hat{F}\,|{\varPsi}_A \rangle |^2 
\delta(\omega+E_\pi-E_K)
\label{eqn:e1}
\end{equation}
with
\begin{equation}
{\hat{F}} 
 =  \int d{\bm r} \> \chi^{(-) \ast}_{{\bm p}_{\pi}}({\bm r})
                     \chi^{(+)}_{{\bm p}_{K}}({\bm r})  
                     \sum_{j=1}^{A}
 ({\overline{f}}+i\,{\overline{g}}{\bm \sigma}_j \cdot{\hat{\bm n}})
                      \hat{\cal O}_{j}\delta ({\bm r}-{\bm r}_{j}),
\label{eqn:e2}
\end{equation}
where $\vert \varPsi_B \rangle$ is a final state of 
the $\Sigma^-$ nuclear system with a total spin $J_B$, 
and $\vert \varPsi_A \rangle$ is an initial state of the 
target nucleus with a total spin $J_A$. 
The momentum and energy transfer to the $\Sigma^-$ final state
is given by 
\begin{equation}
{\bm q}_{\Sigma}={\bm p}_{K}-{\bm p}_\pi, \qquad \omega=E_{K}-E_\pi, 
\label{eqn:e2a}
\end{equation}
where ${\bm p}_{K}$ and ${\bm p}_\pi$ ($E_{K}$ and $E_\pi$) are the 
lab momenta (energies) of the incident $K^-$ and outgoing $\pi^+$ 
in the many-body $K^- + {^A{\rm Z}} \to \pi^+ + {^A_{\Sigma^-}}{\rm (Z-2)}^*$ 
reaction, respectively.
The kinematical factor $\beta$  \cite{Tadokoro1995,Koike2008} expresses 
the translation from the two-body $K^-$--$p$ lab system to 
the $K^-$--${^A{\rm Z}}$ lab system \cite{Dover1983}, which is given by 
\begin{equation}
 \beta=
 \biggl(1+ {E^{(0)}_{\pi} \over E^{(0)}_{\Sigma}}
        {{p^{(0)}_{\pi} - p^{(0)}_{K} \cos\theta_{\rm lab}}
        \over p^{(0)}_{\pi}} \biggr)
        {p_{\pi} E_{\pi} \over p^{(0)}_{\pi} E^{(0)}_{\pi}},
\label{eqn:e3}
\end{equation}
where 
$p^{(0)}_{K}$ and $p^{(0)}_{\pi}$ ($E^{(0)}_{\pi}$ and 
$E^{(0)}_{\Sigma}$) are 
the momenta of $K^-$ and $\pi^+$ (energies of $\pi^+$ and ${\Sigma^-}$)
in the two-body $K^- + p \to \pi^+ + \Sigma^-$ reaction, respectively. 
${\overline{f}}$ and ${\overline{g}}$ in Eq.~(\ref{eqn:e2}) denote 
the non-spin-flip and spin-flip lab amplitudes, respectively,
for the $K^- + p \to \pi^+ + \Sigma^-$ elementary process 
in nuclear medium, 
and the operator ${\hat{\cal O}}_{j}$ changes a $j$th nucleon 
to the $\Sigma^-$ in the nucleus, 
${\hat{\cal O}_j}|\, p_j\, \rangle = |\Sigma^- \rangle$. 
$\chi^{(-)}_{{\bm p}_{\pi}}$ and 
$\chi^{(+)}_{{\bm p}_{K}}$  
are the distorted waves for the outgoing $\pi^+$ and 
incoming $K^-$, respectively. 
The computational procedure for the distorted waves 
is simplified with the help of the eikonal 
approximation \cite{Bouyssy1977,Dover1980}. 
The meson distorted waves are expressed 
\cite{Harada2004} as
\begin{equation}
\chi^{(-)*}_{{\bm p}_{\pi}}({\bm r})
\chi^{(+)}_{{\bm p}_{K}}({\bm r}) 
=\sum_L \sqrt{4 \pi (2L+1)} i^L \tilde{j}_L(q,r)Y_L^0(\hat{\bm r}), 
\label{eqn:e16}
\end{equation}
where  $\tilde{j}_L(q,r)$ is a radial distorted wave with the 
angular-momentum $L$ and momentum transfer $q$.
Here we used total cross sections of $\sigma_{K}$= 32 mb for 
$K^- N$ scattering and $\sigma_\pi$= 30 mb for $\pi^+ N$ one, 
and $\alpha_K =\alpha_\pi=$ 0 \cite{Dover1980}, 
as distortion parameters in $\tilde{j}_L(q,r)$,
together with a matter density distribution fitted to the charge 
radius \cite{deVries1987}.

\Figuretable{TABLE I}

For the nuclear targets, we use single-particle wave functions 
for a proton,
which are calculated with a Woods-Saxon (WS) potential \cite{Bohr1969}:
\begin{equation}
U_N(r)= {V_0^N}f(r)+V^N_{ls} ({\bm l }\cdot{\bm s})
r_0^2 {1 \over r}{d \over dr}f(r)
\label{eqn:e15}
\end{equation}
with $f(r)=[1 + \exp{((r-R)/a)}]^{-1}$, 
where $V^N_{ls}= -0.44 V^N_0$, $a=$ 0.67 fm, 
$r_0=$ 1.27 fm and $R=r_0A^{1/3}$.
We choose the strengths of $V^N_{0}=$ $-$59.7, $-$59.0 and $-$61.7 MeV 
by fits to the charge radii \cite{deVries1987} of 
$\langle r^2_{\rm ch} \rangle^{1/2}=$ 3.09, 3.77 and 5.51 fm 
for $^{28}$Si, $^{58}$Ni and $^{208}$Pb, respectively.
Because deep proton-hole states play an important role 
in describing the $\Sigma^-$ excited and continuum states \cite{Harada2004}, 
we take the single-particle energies and widths 
from ($e,~e'p$) data for nuclei \cite{Jacob1966,Dieperink1990}.
For deep-hole states that are unknown experimentally 
in $^{58}$Ni and $^{208}$Pb, we also use the energies obtained by 
density-dependent Hartree-Fock calculations \cite{Vautherin1972}. 
In Table~\ref{tab:table1}, we list these energies and widths
for $^{28}$Si, $^{58}$Ni and $^{208}$Pb, which are input 
into this calculation.

\subsection{Fermi-averaged $K^-+p\to \pi^+ +\Sigma^-$ cross section}

\begin{figure}[thb]
  \begin{center}
\includegraphics[width=0.7\linewidth]{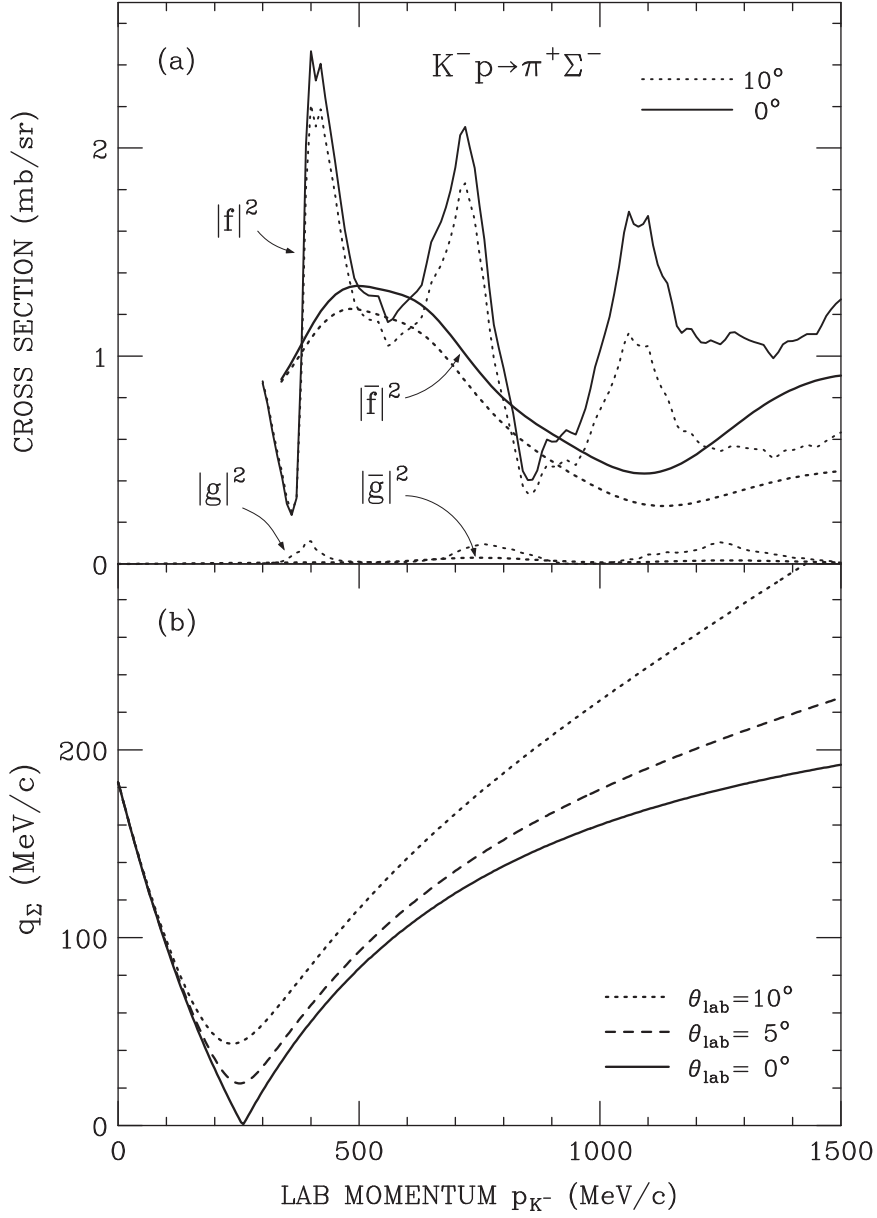}
  \end{center}
  \caption{\label{fig:1}
(a) Absolute cross sections of non-spin-flip and spin-flip processes, 
  $|f|^2$ and $|g|^2$, for the $K^-+p \to \pi^- + \Sigma^-$ reaction 
  in free space \cite{Gopal1977}, 
  as a function of the incident $K^-$ lab momentum $p_{K}$, 
  together with the Fermi-averaged cross sections,
  $|{\overline{f}}|^2$ and $|{\overline{g}}|^2$, in nuclear medium. 
  The solid and dotted curves denote the values at the detected $\pi^+$ angle 
  $\theta_\text{lab}=$ 0$^\circ$ and 10$^\circ$, respectively.
(b) Momentum transfer $q_\Sigma$ for a $\Sigma^-$ production 
  by the ($K^-$,~$\pi^+$) reaction on a $^{58}$Ni target, 
  as a function of the incident $K^-$ lab  momentum $p_{K}$. 
  The solid, dashed and dotted curves
  denote for $\theta_\text{lab}=$ 0$^\circ$, 5$^\circ$ and 10$^\circ$ 
  in the lab frame, respectively.
  }
\end{figure}

When we calculate the nuclear ($K^-$,~$\pi^+$) cross sections 
with the $K^- + p \to \pi^+ + \Sigma^-$ amplitudes, 
it is important to take into account the Fermi-motion 
of a struck nucleon with a Fermi momentum 
$p_F \simeq$ 270 MeV/c in nuclear medium 
\cite{Allardyce1973,Rosenthal1980}.
This effect is considerably enhanced near 
narrow $\Lambda/\Sigma$ resonances 
because their widths are smaller than the Fermi-motion energy 
of the struck nucleon, e.g., $D_{03}$(1520) at $p_{K}\simeq 390$ MeV/c,
$S_{01}$(1670), $D_{03}$(1690) and $D_{13}$(1670) at $\sim 750$ MeV/c, 
as seen in Fig.~\ref{fig:1}(a).
According to the procedure by Rosental and Tabakin \cite{Rosenthal1980}, 
we perform the Fermi-averaging of the 
$K^- + p \to \pi^+ + \Sigma^-$ scattering $T$-matrix 
obtained by Gopal et al. \cite{Gopal1977}.

\Figuretable{FIG. 1}

In Fig.~\ref{fig:1}, we show the Fermi-averaged lab cross 
section of the $K^- + p \to \pi^+ + \Sigma^-$ reaction on nuclei,
\begin{equation}
\left\langle {d \sigma \over d \Omega}
\right\rangle^{K^-p \to \pi^+\Sigma^-}_{\rm lab}
=|\overline{f}|^2+|\overline{g}|^2, 
\end{equation}
at the detected $\pi^+$ angles $\theta_\text{lab}=$ 0$^\circ$ 
and 10$^\circ$, 
as a function of the incident $K^-$ lab momentum $p_{K}$, 
together with the momentum transfer ${q}_{\Sigma}$ in 
the nuclear ($K^-$,~$\pi^+$) reaction. 
$|\overline{f}|^2$ and $|\overline{g}|^2$ 
denote the non-spin-flip and spin-flip components of the 
Fermi-averaged lab cross sections, respectively. 
No data are measured in the low incident $K^-$ 
momentum region below about 300 MeV/c. 
The shape of the Fermi-averaged cross section near 
400--800 MeV/c sizably 
becomes broader, and its value is not so changed by 
a choice of the target, 
as discussed by Dover et al.~\cite{Dover1989,Dover1979}. 
Since the spin-flip cross sections of $|\overline{g}|^2$ 
is negligible, 
we consider only the non-spin-flip process in the 
nuclear ($K^-$,~$\pi^+$) reaction in this paper.

\subsection{Green's function technique}

To evaluate the inclusive spectrum in Eq.~(\ref{eqn:e1}), 
here we employ the Green's function method 
\cite{Morimatsu1988,Morimatsu1994}. 
This technique can describe an unstable hadron nuclear system 
such as a $\Sigma^-$, $\Xi^-$ or $K^-$ nuclear state very 
well \cite{Harada2004,Harada2005,Tadokoro1995,Harada2006a}.
The complete Green's function $G$ provides 
all information concerning $\Sigma$-nucleus dynamics 
as a function of 
the energy transfer $\omega= E_B-E_A$ or 
the energy $E$ measured from the 
$\Sigma^-$ + core-nucleus threshold, 
\begin{equation}
E=E_B-(m_{\Sigma^-}+M_{C})=-B_{\Sigma^-},
\end{equation}
where $m_{\Sigma^-}$ and $M_C$ are masses of 
the $\Sigma^-$ and the core-nucleus, respectively.
It is obtained by solving the 
following potential problems:
\begin{equation}
{G}(E)={G}^{(0)}(E)
+{G}^{(0)}(E)\{{U}_\Sigma+{U}_{\rm Coul}\} {G}(E),
\label{eqn:e12}
\end{equation}
where ${G}^{(0)}(E)$ is a free Green's function. 
${U}_\Sigma$ is the $\Sigma^-$-nucleus potential, 
and ${U}_{\rm Coul}$ is the finite Coulomb potential between the 
$\Sigma^-$ and the core-nucleus.
By a use of the complete Green's function ${G}$, 
a sum of the final states $B$ of Eq.~(\ref{eqn:e1}) is given as
\begin{equation}
 \sum_{B}  \vert \varPsi_B \rangle \delta (\omega+E_{A}-E_{B})
    \langle \varPsi_B \vert 
  = -{1 \over \pi} {\rm Im}{G}(\omega).
\label{eqn:e24}
\end{equation}
Thus the inclusive spectrum of the double-differential cross section 
is rewritten as
\begin{equation}
  {{d^2\sigma} 
 \over {d\Omega_{\pi} dE_{\pi}} }
 =  {\beta}
\left\langle {d \sigma \over d \Omega}
\right\rangle^{K^-p \to \pi^+\Sigma^-}_{\rm lab}
S(\omega)
 \label{eqn:e25}
\end{equation}
with the strength function $S(\omega)$, which is given by
\begin{equation}
S(\omega)=
-{1 \over \pi}{\rm Im} \sum_{\alpha' \alpha} 
 \int\,d{\bm r}' d{\bm r} f_{\alpha'}^\dagger({\bm r}') 
   G_{\alpha' \alpha}(\omega;{\bm r}',{\bm r})
  f_{\alpha}({\bm r}),
 \label{eqn:e25a}
\end{equation}
where ${\bm r}$ is the relative coordinate between the $\Sigma^-$ and 
the core-nucleus. 
$f_{\alpha}({\bm r}) $ presents the production function via
$\Sigma^-$-nucleus doorways that are excited initially as
\begin{equation}
f_{\alpha}({\bm r}) = 
\chi^{(-)*}_{{\bm p}_{\pi}}({M_{C} \over M_{B}}{\bm r})
\chi^{(+)}_{{\bm p}_{K}}({M_{C} \over M_{A}}{\bm r})
\langle \alpha | \hat{\psi}_N ({\bm r}) | \varPsi_{A} \rangle,
\label{eqn:e26}
\end{equation}
where
$\langle \alpha \, | \hat{\psi}_N ({\bm r}) |\varPsi_{A} \rangle$ 
is a hole-state wave function for a struck nucleon in the 
target, and $\alpha$ denotes the complete set of eigenstates 
for the system.
The factors of $M_{C}/M_{B}$ and $M_{C}/M_{A}$ 
take into account the recoil effects.

Here we consider the $\Sigma^-$-nucleus system in a 
non-relativistic framework. Since $U_\Sigma$ has an imaginary part, 
the Hamiltonian ${\mathcal H}=T+U_{\Sigma}$ 
is non-Hermitian. Thus the Schr\"odinger equations are written as  
\begin{equation}
{\mathcal H} \, {\varphi}_{n} = E_{n} \, {\varphi}_{n}, 
\qquad 
{\mathcal H}^\dagger \, \tilde{\varphi}_{n} 
= E^*_{n} \, \tilde{\varphi}_{n}, 
\label{eqn:e57c}
\end{equation}
where 
$E_{n}$ is a \textit{complex} eigenvalue.
The $\Sigma^-$ nuclear binding energy and width
for a $\Sigma^-$ unstable bound state
can be evaluated as 
\begin{eqnarray}
E_{n}={(k^\text{(pole)}_{n})^2/2 \mu}=-B_{n}-i{\mit\Gamma}_{n}/2,
\label{eqn:e57a}
\end{eqnarray}
where $k^\text{(pole)}_{n}$ denotes a pole position of 
the bound state in the complex $k$-plane 
(${\rm Re} k^\text{(pole)}_{n} < 0$, ${\rm Im} k^\text{(pole)}_{n} > 0$),
and $\mu$ is the reduced mass of the $\Sigma^-$-nucleus system. 
${\varphi}_{n}$ is a wave function of the eigenstate labeled by 
${k^\text{(pole)}_{n}}$
and $\tilde{\varphi}_{n}$ is the wave function given by a 
\textit{biorthogonal} set;
its conjugate state becomes $(\tilde{\varphi}_{n})^*={\varphi}_{n}$, 
of which radial wave functions must be normalized by 
so-called $c$-products \cite{Berggren1968},  
\begin{eqnarray}
&&\int_0^{\infty} r^2d{r} 
(\tilde{\varphi}_{n}(r))^*{\varphi}_{n}(r) \nonumber\\
&&=\int_0^{\infty} r^2d{r} ({\varphi}_{n}(r))^2=1, 
\label{eqn:e56c}
\end{eqnarray}
under the boundary condition for decaying states \cite{Kapur1938}. 
The completeness relation for the complete Green's function 
is written as
\begin{eqnarray}
G(E;{r}',{r})
&=&
\sum_n {\varphi_{n}(r')(\tilde{\varphi}_{n}(r))^* \over 
E-E_{n}+i\epsilon} \nonumber \\
&+& 
{2 \over \pi}\int_0^\infty dk {k^2 S(k) 
u(k,r')(\tilde{u}(k,r))^* \over E-E_k+i\epsilon},
\label{eqn:e57d}
\end{eqnarray}
where the summation over $n$ includes all the 
pole of the $S$-matrix in the complex $k$-plane, 
and $u(k,r)$ is a scattering wave function. 
In the bound region, 
Green's function might be expanded \cite{Morimatsu1994} as   
\begin{eqnarray}
G(E;{r}',{r})
&=&
\sum_n G_{n}^\text{(pole)}(E;{r}',{r}) + G^\text{(bg)}(E;{r}',{r}),
\label{eqn:b1}
\end{eqnarray}
where the pole contribution of the bound state 
can be expressed as 
\begin{eqnarray}
G_{n}^\text{(pole)}(E;{r}',{r})
= {\varphi_{n}(r')(\tilde{\varphi}_{n}(r))^* 
\over E-E_{n}+i\epsilon},
\label{eqn:b2}
\end{eqnarray}
and 
$G^\text{(bg)}(E;{r}',{r})$ indicates the background contribution.
Note that the interference by the background term or other pole 
terms occasionally affects the spectrum in the continuum region 
$E > 0$ near the $\Sigma^-$ threshold if $|{\rm Im}U_\Sigma|$ 
is large \cite{Morimatsu1994}.

\subsection{Integrated cross sections and the complex effective number}

To study a structure of hypernuclear bound states, the 
integrated cross sections of these states have been often evaluated in 
DWIA calculations
\cite{Hufner1974,Bouyssy1977,Dover1980,Auerbach1983,Dover1983,Motoba1988}.
The angular distributions of the cross section are presented 
as a function of $\theta_{\rm lab}$ or $q_\Sigma$.
The integrated cross section of 
the $\Sigma^-$ unstable bound state with $(j^{-1}_pj_{\Sigma})J_B$
in the nuclear ($K^-$,~$\pi^+$) reaction is written as 
(see Appendix B in Ref.~\cite{Koike2008}):
\begin{eqnarray}
\biggl({d\sigma~ \over d{\Omega_\pi}}\biggr)
&=& \int {dE_\pi}\, 
\biggl({d^2\sigma \over dE_\pi d{\Omega_\pi}}\biggr) \nonumber\\
&=&  
\bar{\alpha} \left\langle{d\sigma \over d\Omega} 
\right\rangle_{\rm lab}^{K^- p \to \pi^+ \Sigma^-}\!\!\!
{\rm Re}\, P^{(j^{-1}_pj_{\Sigma})J_B}_{\rm eff},
\label{eqn:e52}
\end{eqnarray}
where 
$\bar{\alpha}$ is a kinematical factor 
\cite{Dover1980,Dover1983,Motoba1988,Tadokoro1995} 
defined by
\begin{equation}
 \bar{\alpha} = \beta
{\biggl(1+ {E_\pi \over E_B}{{p_\pi - p_K \cos\theta_{\rm lab}} 
       \over p_\pi} \biggr)^{-1}}, 
\label{eqn:b11}
\end{equation}
and $P_{\rm eff}$ is
the \textit{complex} effective number of a proton \cite{Koike2008},
which describes all information on the nuclear structure of 
the unstable bound systems.
When the potential has no imaginary part, 
$P_{\rm eff}$ becomes a real number.
This approach also provides a good insight into 
a signal of $\Sigma^-$ bound states caused by 
a complex potential $U_\Sigma$, 
as discussed in Refs.~\cite{Morimatsu1994,Koike2008}.

For a closed-shell target with $J^\pi=$ 0$^+$, 
the complex effective number of a proton 
for the transition $J_A \to {(j^{-1}_p j_{\Sigma})J_B}$ 
is written as 
\begin{eqnarray}
P^{(j^{-1}_p j_{\Sigma})J_B}_{\rm eff}
&=& (2J_B+1)(2 j_{\Sigma}+1)(2j_p+1) \nonumber\\
&& \times   \left( \begin{array}{ccc}
     j_{\Sigma} & j_p & J_B  \\
     {1 \over 2}  & -{1 \over 2} & 0
    \end{array} \right)^2
    F({q})F^\dagger({q}),
\label{eqn:e54}
\end{eqnarray}
where $\ell_{\Sigma}+\ell_p+J_B$ must 
be even because the spin-flip processes are neglected.
This leads to a population of the natural parity 
states with $J^\pi=$ 0$^+$, 1$^-$, $2^+$, $\cdots$.
The form factor $F(q)$ is
\begin{eqnarray}
F(q)&=&
\int_0^{\infty} r^2d{r} 
(\tilde{\varphi}_{j_{\Sigma}}(r))^*
\tilde{j}_{J_B}(q,br)\varphi^{(N)}_{j_p}(r), 
\label{eqn:e55}
\end{eqnarray}
where
$\varphi^{(N)}_{j_p}$ is a single-particle 
wave function for the proton, and 
$\tilde{\varphi}_{j_{\Sigma}}$ is the biorthogonal one 
for $\Sigma^-$, as given by Eq.~(\ref{eqn:e57c}). 
The recoil effects are taken into account in the distorted waves of 
$\tilde{j}_{J_B}(q,br)$
by the factor of 
$b= {M_{C}/M_{B}}$ when ${M_{B} \simeq M_{A}}$ 
is justified in Eq.~(\ref{eqn:e26}).

\section{$\Sigma$-nucleus potentials \label{sect:pot}}

We briefly mention the $\Sigma$-nucleus potentials
which we discussed in this paper.
In previous papers \cite{Harada2005,Harada2006a}, we have introduced 
several types of the $\Sigma$-nucleus potential
obtained by fitting to strong-interaction shifts and 
widths of $\Sigma^-$ atomic X-ray data; 
(a) the density-dependent (DD) potential \cite{Batty1994},
(b) the relativistic mean-field (RMF) potential \cite{Mares1995},
(c) the local-density approximation potential (LDA-NF) based on 
YNG-NF interaction \cite{Yamamoto1985,Dabrowski2002},
(d) the LDA potential (LDA-S3) based on phenomenological 
two-body $\Sigma N$ SAP-3 interaction \cite{Harada1995b},
(e) the shallow potential in the WS form (WS-sh) \cite{Hayano1988}, 
and (f) the $t_{\rm eff}\rho$-type potential ($t_{\rm eff}\rho$)
\cite{Batty1994}.
It should be noticed that all of the potentials sufficiently 
reproduce the experimental shifts and 
widths of the $\Sigma^-$ atomic states;
the values of the shifts and widths are mainly sensitive 
to the tail part of the potentials outside 
the nuclear surface \cite{Batty1994}. 

\Figuretable{FIG. 2}

\begin{figure}[htb]
\begin{center}
\includegraphics[angle=90,width=0.7\linewidth]{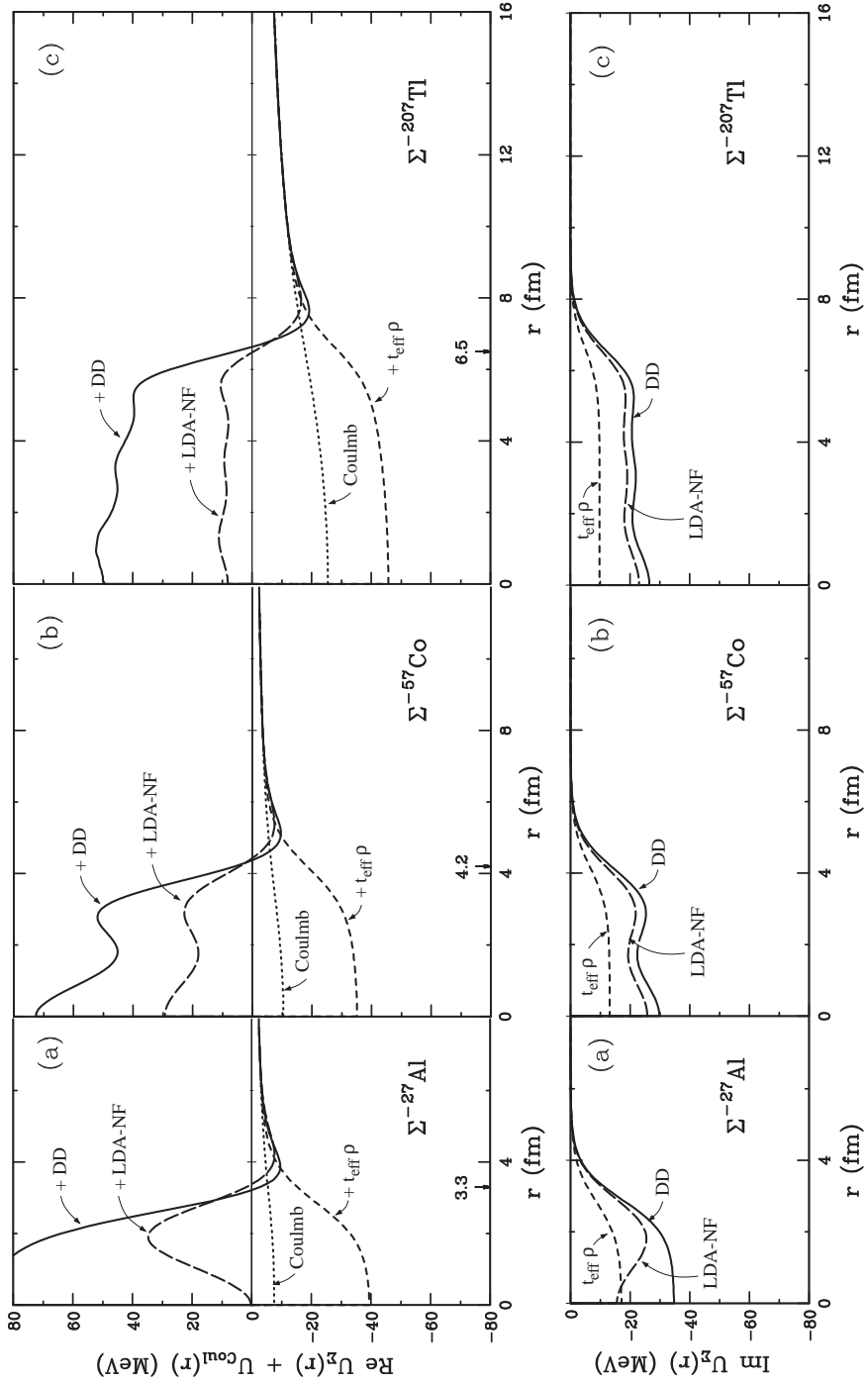}
  \caption{\label{fig:2}
  (top) Real and (bottom) imaginary parts of the $\Sigma$-nucleus potential
  plus the finite Coulomb potential 
  for (a) $\Sigma^-$--$^{27}$Al, (b) $\Sigma^-$--$^{57}$Co and (c)
  $\Sigma^-$--$^{207}$Tl. 
  The solid, long-dashed and dashed curves denote the radial distribution of 
  the potentials for DD, LDA-NF and $t_{\rm eff}\rho$, respectively.
  The strength for the real part includes the finite Coulomb potential.
  The dotted curves denote only the Coulomb potential for the 
  $\Sigma^-$-nucleus systems.
  }
  \end{center}
\end{figure}

In Fig.~\ref{fig:2}, we display the real and imaginary parts of 
the $\Sigma$-nucleus potentials of DD, LDA-NF, 
and $t_{\rm eff} \rho$ 
for $\Sigma^-$--$^{27}\text{Al}$ ($^{28}_{\Sigma^-}$Mg), 
$\Sigma^-$--$^{57}\text{Co}$ ($^{58}_{\Sigma^-}$Fe) 
and $\Sigma^-$--$^{207}\text{Tl}$ ($^{208}_{\Sigma^-}$Hg),
where 
the real parts of the potential include 
the finite Coulomb potentials.
In Refs.~\cite{Harada2005,Harada2006a}, 
it has been attempted to discriminate between 
these types of the potentials by analyzing 
the nuclear ($\pi^-$,~$K^+$) data in the 
$\Sigma^-$ continuum spectrum. 
These analyses have shown that the $\Sigma$-nucleus potentials 
have a repulsion inside the nuclear surface and an attraction outside 
the nucleus, i.e., DD, RMF and LDA-NF, 
rather than an attraction at the nuclear center, 
i.e., LDA-S3, WS-sh and $t_\text{eff} \rho$.
Moreover, the former potentials are considerably different 
from each other 
in terms of the repulsion at $r \lesssim$ $R=r_0(A-1)^{1/3}$ fm and the 
attractive pocket outside there.
The repulsion in DD at the nuclear center (${\sim}80$ MeV)
is 2--3 times larger than that in LDA-NF or RMF (${\sim}40$ MeV)
\cite{Harada2005},
whereas the range of the pockets in DD near the nuclear surface 
is slightly shorter than that in LDA-NF.  
Such a difference of the repulsion and attractive pocket is 
expected to be clearly examined if we can observe 
the $\Sigma^-$ bound states in the spectrum. 
In this paper we will focus on three potentials of 
DD, LDA-NF and $t_{\rm eff} \rho$, as the typical examples.

\section{${\bm \Sigma^-}$-nucleus bound states \label{sect:bound}}

\subsection{Energies and widths}

In Table~\ref{tab:table2}, we show the numerical 
results of the 
$\Sigma^-$ nuclear binding energies and widths of the bound states 
for $\Sigma^-$--$^{27}\text{Al}$, $\Sigma^-$--$^{57}\text{Co}$ 
and $\Sigma^-$--$^{207}\text{Tl}$,  
when we use the DD or $t_{\rm eff}\rho$ potential 
including the finite Coulomb potential. 

\Figuretable{TABLE II}
\Figuretable{FIG. 3}

\begin{figure}[bth]
  \begin{center}
  \includegraphics[width=0.9\linewidth]{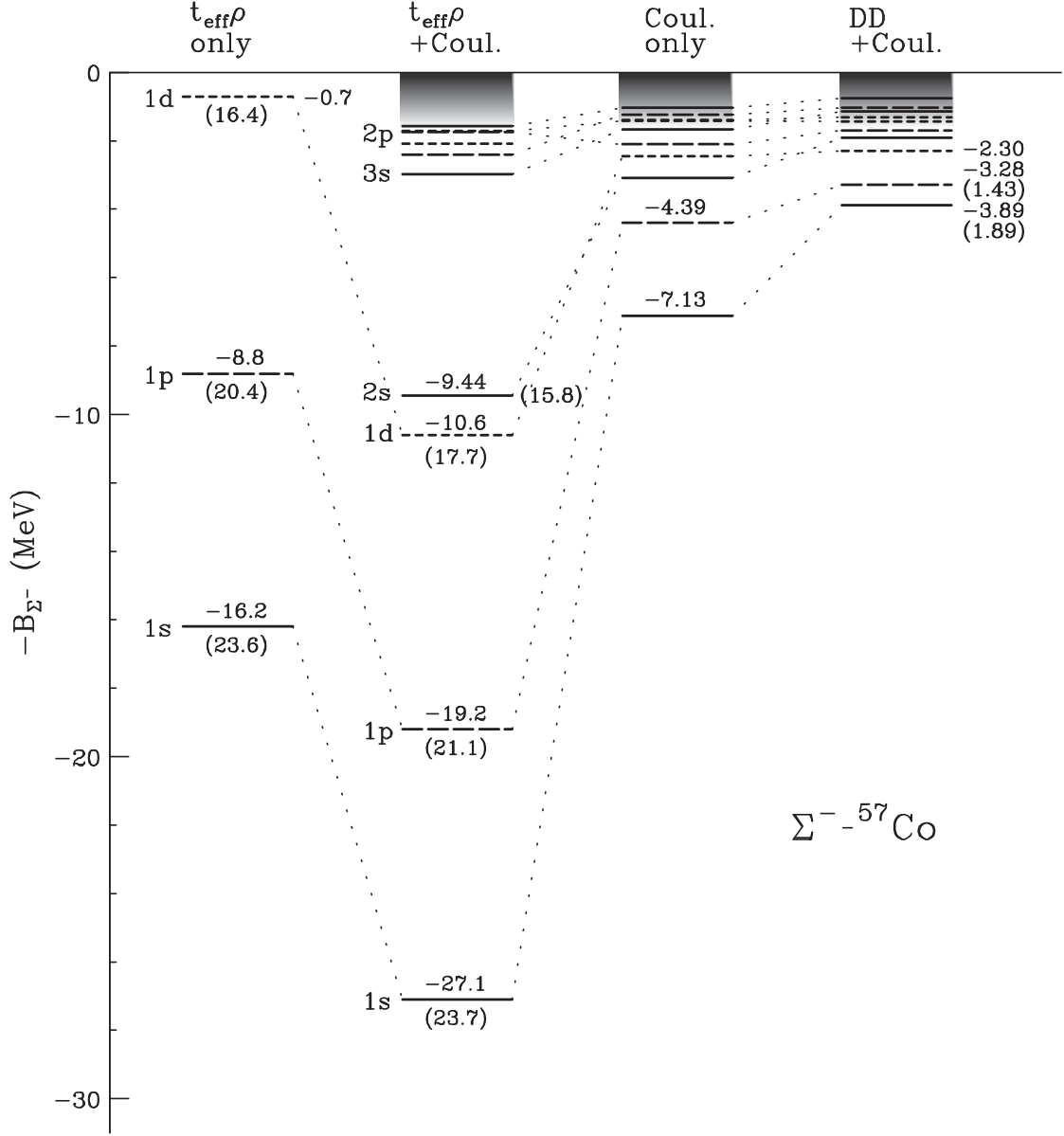}
  \end{center}
  \caption{\label{fig:3}
  Binding energies and widths (in the brackets) of the $\Sigma^-$ $(ns)_\Sigma$, 
  $(np)_\Sigma$ and $(nd)_\Sigma$ 
  bound states in $\Sigma^-$--$^{57}$Co.
  These values are obtained by the $t_{\rm eff}\rho$ potential only, 
  the $t_{\rm eff}\rho$ or DD potential
  including the finite Coulomb potential, and the finite Coulomb 
  potential only. 
  The $\Sigma^-$ atomic states occur in the shaded region.
  See Table~\ref{tab:table2}.
  }
\end{figure}

For the $t_{\rm eff}\rho$ potential, we confirm that
there exist $\Sigma^-$ bound states 
even if the Coulomb potential is switched off (see Fig.~\ref{fig:3}), 
because its real part corresponds 
to $V_0^\Sigma \simeq$ $-$28 MeV as a WS potential,
which is similar to the $\Lambda$-nucleus 
potential \cite{Dover1980,Motoba1988}. 
Such a bound state has a broad width 
($\varGamma_\Sigma/2 \agt B_{\Sigma^-}$), 
of which the pole arises away from the physical 
axis in the complex $k$-plane \cite{Gal1981}. 
If the pole is located close to the $\Sigma^-$ emitted threshold, 
the magnitude of the production peak is modified 
by interference effects from background or other pole terms 
\cite{Morimatsu1988}. 
Thus the shape of the peak in the spectrum does not 
necessarily correspond to that of a standard Breit-Wigner 
resonance located at $B_{\Sigma^-}$ \cite{Morimatsu1988,Koike2008}.

For the DD potential, on the other hand, we do not 
obtain the bound states till we take into account 
the Coulomb attraction 
because the potential is very repulsive inside the nucleus. 
As seen in Table~\ref{tab:table2}, the $\Sigma^-$ binding 
energies of $(1s)_\Sigma$, $(1p)_\Sigma$ and $(1d)_\Sigma$ 
states are significantly shifted upward 
from those of the finite Coulomb eigenstates (see Fig.~\ref{fig:3}), 
and their widths become rather narrow (${\sim}1$ MeV). 
A relatively narrow width is obtained 
in the $\Sigma^-$ bound states because of the repulsion 
even if the value of $|{\rm Im}U_{\Sigma}|$ becomes larger.
It is recognized that these states are regarded as 
{\it Coulomb-assisted $\Sigma^-$-nucleus bound states}.  
In Fig.~\ref{fig:3}, we illustrate a summary of 
the $\Sigma^-$ nuclear binding energies and widths for 
low-lying bound states of $\Sigma^-$--${^{57}{\rm Co}}$.
The higher excited states look more like $\Sigma^-$ atomic 
states and the level density of $\Sigma^-$ increases toward the 
$\Sigma^-$ threshold.

\subsection{Wave functions}

\Figuretable{FIG. 4}

\begin{figure}[htb]
  \begin{center}
\includegraphics[angle=90,width=0.68\linewidth]{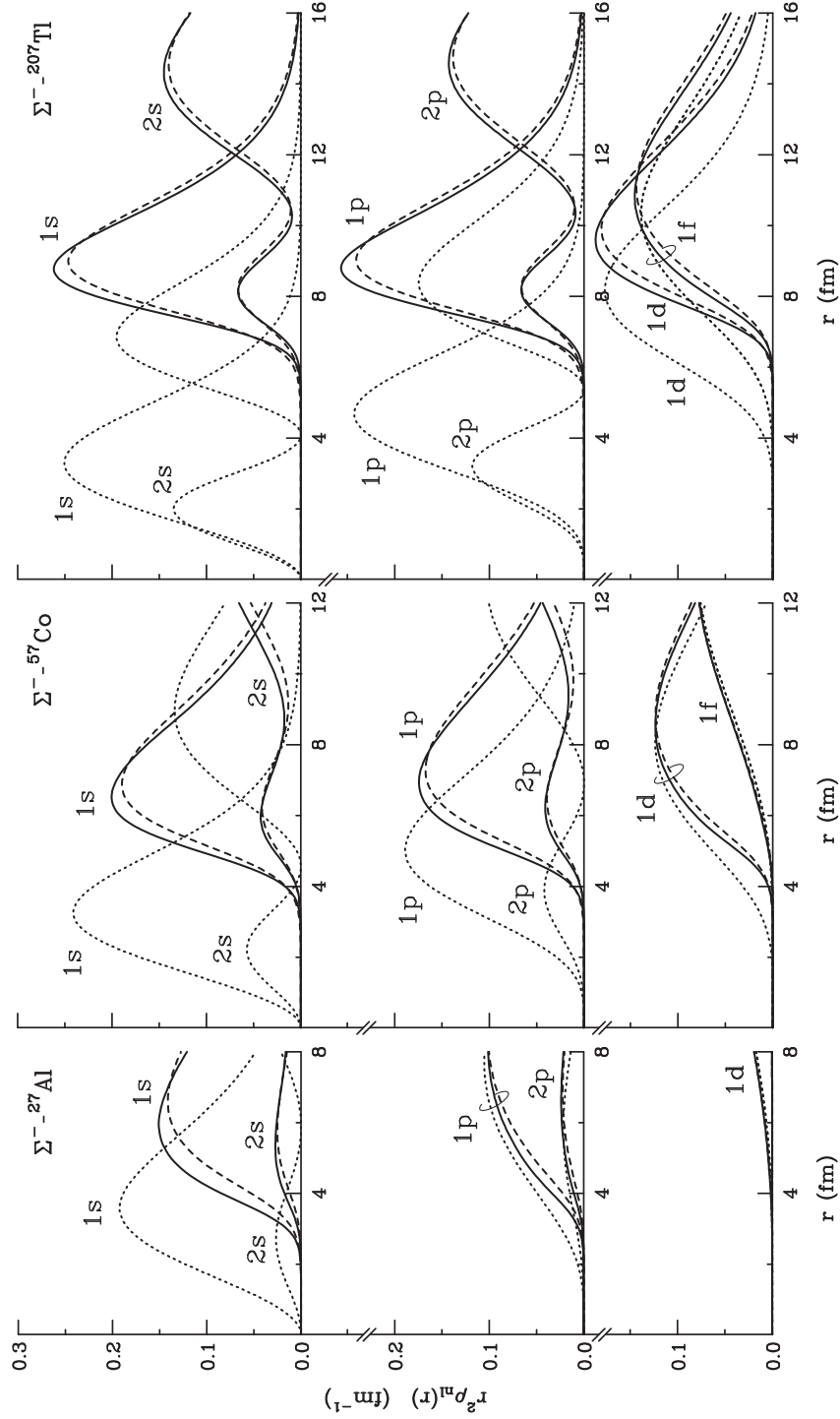}
\end{center}  
  \caption{\label{fig:4}
  Density distributions of $\Sigma^-$-nucleus $1s$, $2s$, $1p$,
  $2p$, $1d$ and $1f$ bound states in $\Sigma^-$--$^{27}$Al (left),  
  $\Sigma^-$--$^{57}$Co (middle) and $\Sigma^-$--$^{207}$Tl (right) systems. 
  The solid and dashed curves denote the density distributions of 
  $r^2 \rho_{n\ell}(r)$ for the DD and LDA-NF potentials, respectively, 
  taken into account the finite Coulomb potential. 
  The dotted curves denote the density distributions of the finite Coulomb 
  bound states.  
  }
\end{figure}

To clearly understand the effects of the repulsion 
such as the DD potential at the nuclear inside, 
we examine behavior of wave functions of 
the $\Sigma^-$ bound states.
In Fig.~\ref{fig:4}, we display the density distributions of
$r^2 \rho_{n\ell}(r) = r^2|\varphi_{n\ell}(r)|^2$
that are calculated with the DD or LDA-NF potential, 
for $(1s)_\Sigma$, $(2s)_\Sigma$, $(1p)_\Sigma$, $(2p)_\Sigma$, 
$(1d)_\Sigma$, $(2d)_\Sigma$ and $(1f)_\Sigma$ states 
in $\Sigma^-$--$^{27}$Al, $\Sigma^-$--$^{57}$Co and 
$\Sigma^-$--$^{207}$Tl.
The wave functions of the 
$(ns)_\Sigma$, $(np)_\Sigma$ and $(nd)_\Sigma$ states 
for both DD and LDA-NF are significantly pushed out 
to the nuclear outside because of the strong repulsion 
inside the nucleus, 
in comparison with the Coulomb wave functions.
The overlap between wave functions and the 
imaginary part of the potential is small, so that 
their widths are considerably reduced. 
For $\Sigma^-$--$^{27}$Al, indeed, the $(1s)_\Sigma$ state has a 
narrow width of $\varGamma_{\Sigma} =$ 1.25 MeV and 
a rms radius of $\langle r^2 \rangle^{1/2}_{1s}$= 9.8 fm. 
For $\Sigma^-$--$^{57}$Co, the $(1s)_\Sigma$ state has 
$\varGamma_{\Sigma} =$ 1.89 MeV and 
$\langle r^2 \rangle^{1/2}_{1s}$= 9.1 fm, 
and the $(1p)_\Sigma$ state also $\varGamma_{\Sigma} =$ 1.43 MeV and 
$\langle r^2 \rangle^{1/2}_{1s}$= 9.8 fm.
If the DD or LDA-NF potential is switched off, 
we have a rms radius of $\langle r^2 \rangle^{1/2}_{1s}=$ 4.2 fm
for the $\Sigma^-$ atomic $(1s)_\Sigma$ state 
in $\Sigma^-$--$^{57}$Co, 
of which the core-nuclear radius is $R= r_0 A^{1/3}=$ 4.23 fm.
The $\Sigma^-$ hyperon in nuclei is excluded from 
the nuclear region, and it acts on a \textit{strangeness halo} 
in the Coulomb-assisted $\Sigma^-$-nucleus bound states.  

This nature becomes more conspicuous in heavy nuclei. 
For $\Sigma^-$--$^{207}$Tl, 
wave functions for these states with DD and LDA-NF
are extremely pushed out to the nuclear outside 
because of the strong repulsion inside the nucleus.
The binding energy and a rms radius of the $(1s)_{\Sigma}$ 
state for DD are $B_{\Sigma^-}=$ $-$10.5 MeV and 
$\langle r^2 \rangle^{1/2}_{1s}$= 9.8 fm, 
respectively, in contrast to $B_{\Sigma^-}=$ $-$18.8 MeV and 
$\langle r^2 \rangle^{1/2}_{1s}$= 4.2 fm 
in only the Coulomb $(1s)_\Sigma$ state.
If the $\Sigma^-$ narrow bound state with a width 
of $\varGamma_{\Sigma}$= 1.7 MeV 
can be selectively populated, 
one expects to clearly see a distinct peak of the state 
in the $\Sigma^-$ bound region.
For details of the repulsive component in the $\Sigma$-nucleus 
potential, 
moreover, it is important to find a quantitative different signal
between DD and LDA-NF in the calculated spectrum of the 
$\Sigma^-$ bound states. As shown in Fig.~\ref{fig:4}, 
however, a discrepancy between these wave functions 
for $\Sigma^-$--$^{207}$Tl is rather small 
because these wave functions 
are predominately located near the attractive pocket formed 
by the Coulomb potential.
Consequently, it might be difficult to distinguish the 
repulsion inside the nuclei between the DD and LDA-NF potentials for 
heavy nuclei such as $\Sigma^-$--$^{207}$Tl systems.
The production spectrum for the ($K^-$, $\pi^+$) reaction on a 
$^{208}$Pb target will be discussed 
in Sect.~\ref{sect:208Pb}.

\section{\label{sect:results}
Results and Discussion}

As mentioned in Sect.~\ref{sect1},  
the ($K^-$,~$\pi^+$) reaction provides the ability 
of a production of ``substitutional states'' 
under the recoilless condition, 
in contrast to the exothermic ($\pi^-$,~$K^+$) reactions.
Because the $K^- + p \to \pi^+ + \Sigma^-$ spin-flip process
is negligible in this momentum region, 
the ($K^-$,~$\pi^+$) reaction on the closed-shell 
nuclear targets with $J^\pi=$ 0$^+$ 
can selectively populate natural-parity states with 
$J^\pi=$ 0$^+$, 1$^-$, $2^+$, $\cdots$, as seen in Eq.~(\ref{eqn:e54}),
in only the isospin transfer $\Delta T=$ 3/2. 
Thus this reaction is advantageous to populate the 
Coulomb-assisted $\Sigma^-$ bound states in the nucleus \cite{Tadokoro1991}.
We expect to extract more quantitative information 
on a repulsive component of the $\Sigma$-nucleus potential 
from the nuclear ($K^-$,~$\pi^+$) spectrum,  
by choosing an appropriate target nucleus and kinematics.
Now we examine the production spectra of the 
Coulomb-assisted $\Sigma^-$-nucleus bound states by the 
($K^-$,~$\pi^+$) reactions on $^{28}$Si, $^{58}$Ni, 
and $^{208}$Pb targets.
In this paper, we assume a detector resolution 
of 1.5 MeV FWHM for the following calculated spectra.

\subsection{$^{58}$Ni($K^-$,~$\pi^+$) spectrum
\label{sect:58Ni}}

Let us consider the Coulomb-assisted $\Sigma^-$-nucleus bound states 
by the ($K^-$,~$\pi^+$) reaction on the $^{58}$Ni target. 
The nucleus $^{58}$Ni is very suitable because it is 
nearly a subclosed-shell nucleus of the proton $f_{7/2}$ orbit, 
and its proton-hole strength is well 
concentrated on the $J^\pi={7/2}^-$,~$T={3/2}$
ground state of $^{57}$Co \cite{Marinov1985}.
Thus the $^{58}_{\Sigma^-}$Fe hypernucleus, which consists 
of the $\Sigma^-$ and the $^{57}$Co nucleus, can be produced 
with total isospin $T={5/2}$, $T_z=-{5/2}$.
As seen in Fig.~\ref{fig:3}, the energies and widths of 
the $\Sigma^-$--$^{58}{\rm Co}$ bound states 
with the DD potential are very different from those 
with the $t_{\rm eff}\rho$ potential, 
as well as those with only the finite Coulomb one. 
To establish properties of the $\Sigma$-nucleus 
potentials for $\Sigma^-$--$^{57}$Co, 
we demonstrate the production spectra of the 
$\Sigma^-$-nucleus bound states by utilizing the 
near-recoilless ($K^-$,~$\pi^+$) reaction.

\subsubsection{Strength function}

\begin{figure}[htb]
  \begin{center}
  \includegraphics[width=0.9\linewidth]{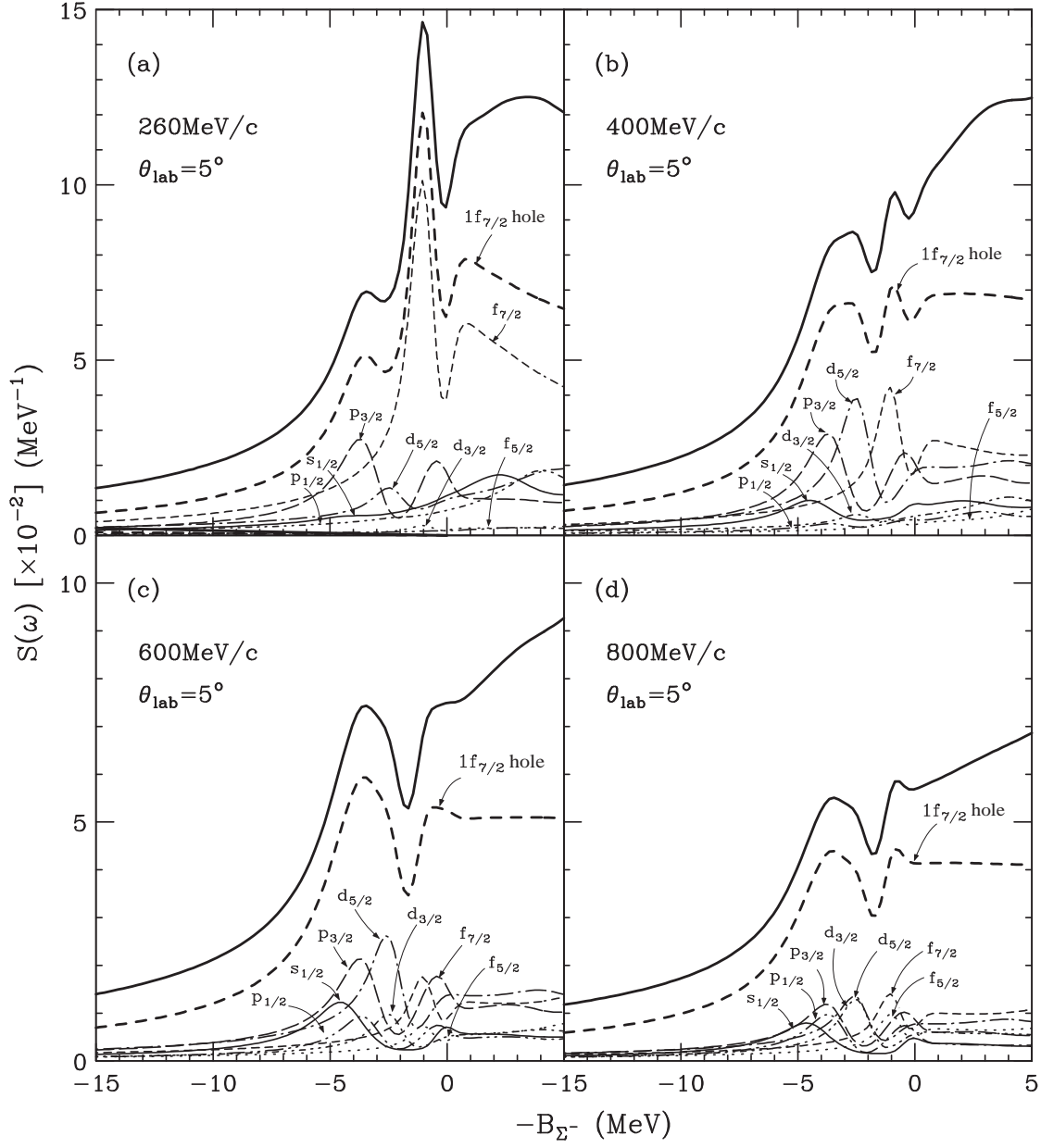}
  \end{center}
  \caption{\label{fig:5}
  Partial-wave components of the calculated strength function $S(\omega)$ 
  for the inclusive spectra of the $^{58}$Ni($K^-$,~$\pi^+$) reaction 
  near the $\Sigma^-$ threshold
  at (a) $p_K=$ 260 MeV/c (5$^\circ$), (b) 400 MeV/c (5$^\circ$), 
  (c) 600 MeV/c (5$^\circ$) and (d) 800 MeV/c (5$^\circ$), as a function 
  of a $\Sigma^-$ binding energy $-B_{\Sigma^-}$. 
  Here the DD potential is used. 
  The thick solid and thick dashed curves denote a total inclusive spectrum 
  and a total contribution of a $1f_{7/2}^{-1}$ proton-hole state 
  in the $^{58}$Ni target. 
  The spectra are folded with a detector resolution of 1.5 MeV FWHM. 
  }
\end{figure}

By the Green's function technique, we calculate 
the inclusive spectra of the $\Sigma^-$--$^{57}$Co 
bound states with the DD potential
in the $^{58}$Ni($K^-$,~$\pi^+$) reaction. 
To clarify the dependence of the spectrum on the momentum 
transfer $q_\Sigma$, 
we evaluate the strength function of $S(\omega)$ in Eq.~(\ref{eqn:e25a}).
In Fig.~\ref{fig:5}, we show the strength functions $S(\omega)$ 
near the $\Sigma^-$ threshold on the incident 
$K^-$ lab momenta of (a) $p_{K}=$ 260 MeV/c, 
(b) 400 MeV/c, (c) 600 MeV/c and (d) 800 MeV/c 
with the angle $\theta_{\rm lab}=5^\circ$, 
which correspond to the momentum transfers of $q_{\Sigma}=$ 22 MeV/c, 66 MeV/c, 
117 MeV/c and 153 MeV/c, respectively (see Fig.~\ref{fig:1}(b)).
The kinematical factor $\beta$ is taken to be $\beta=$ 1.02--0.89 
depending on $-B_{\Sigma^-}=$ $(-30)$--$(+20)$ MeV at 600 MeV/c (5$^{\circ}$). 
The inclusive cross sections can be obtained by Eq.~(\ref{eqn:e25}).

\Figuretable{FIG.~5}

For $p_{K}=$ 260 MeV/c ($q_{\Sigma}=$ 22 MeV/c), 
which leads to an almost recoilless kinematics, 
we find that a series of $f_{\Sigma}$ states involving 
the higher-excited $\Sigma^-$ {\it atomic} bound and continuum 
states with a couple to an $f_{7/2}^{-1}$ proton-hole state is strongly favored 
because a $\Delta L=$ 0 transition is dominant, 
as well as a substitutional $(1f^{-1}_{7/2},1f_{\Sigma})_{0^+}$ state
at $B_{\Sigma^-}=$ 1.43 MeV. 
Such a spectrum is rather unsuitable to see the $\Sigma^-$ 
{\it nuclear} bound states to extract the repulsive component of 
the $\Sigma$-nucleus potential. 
The value of 
$\langle d\sigma/d\Omega \rangle^{K^-p\to \pi^+\Sigma^-}_{\rm lab}$ 
is experimentally unknown (see Fig.~\ref{fig:1}(a)), and 
a corresponding measurement seems to be difficult due to 
a background of $K^- \to \pi^+ \pi^- \pi^-$ decays 
in the low momentum $K^-$ beams at the forward direction.
For $p_K=$ 400 MeV/c ($q_\Sigma\simeq$ 66 MeV/c) where 
$\langle d\sigma/d\Omega \rangle^{K^-p\to \pi^+\Sigma^-}_{\rm lab}=$ 
1.13 mb/sr, 
we find that the partial-wave components of 
$(s_{1/2})_{\Sigma}$, $(p_{3/2})_{\Sigma}$,
$(d_{5/2})_{\Sigma}$ and $(f_{7/2})_{\Sigma}$
are fairly populated with a couple to an $f_{7/2}^{-1}$ 
proton-hole state, leading to a bump with 
$-B_{\Sigma^-} \simeq -3.5$ MeV and 
$\varGamma_\Sigma \simeq$ 4MeV. 
Although $q_\Sigma\simeq$ 66 MeV/c is still small,  
the peak for the $(f_{7/2})_{\Sigma}$ state is reduced 
and the bump structure is formed in the spectrum. 
For $p_K=$ 600 MeV/c ($q_\Sigma \simeq$ 117 MeV/c)
where $\langle d\sigma/d\Omega \rangle^{K^-p\to \pi^+\Sigma^-}_{\rm lab}
=$ 1.25 mb/sr is near its maximum, 
we show that the bump structure is mainly constructed with 
partial-wave components of 
$(s_{1/2})_\Sigma$, $(p_{3/2})_\Sigma$ and $(d_{5/2})_\Sigma$, 
but the contribution of the $(f_{7/2})_\Sigma$ component is small.
Thus the bump structure appears more clearly below the $\Sigma^-$ 
threshold in the spectrum.
For $p_K=$ 800 MeV/c ($q_{\Sigma}\simeq$ 153 MeV/c), 
we find that the contributions of $(d_{5/2})_{\Sigma}$ and 
$(d_{3/2})_{\Sigma}$ to the bump are comparable, and the 
dip around 
$-B_{\Sigma^-}\simeq$ $-$2 MeV tends to be filled by other 
partial-wave contributions. 
Consequently, the momentum transfer of the near-recoilless 
$q_\Sigma \simeq 110$ MeV/c at $p_K \simeq 600$ MeV/c
is necessary to clearly obtain the bump structure of the 
Coulomb-assisted $\Sigma^-$ bound states in the spectrum.

\Figuretable{FIG.~6}

\begin{figure}[htb]
  \begin{center}
  \includegraphics[width=0.9\linewidth]{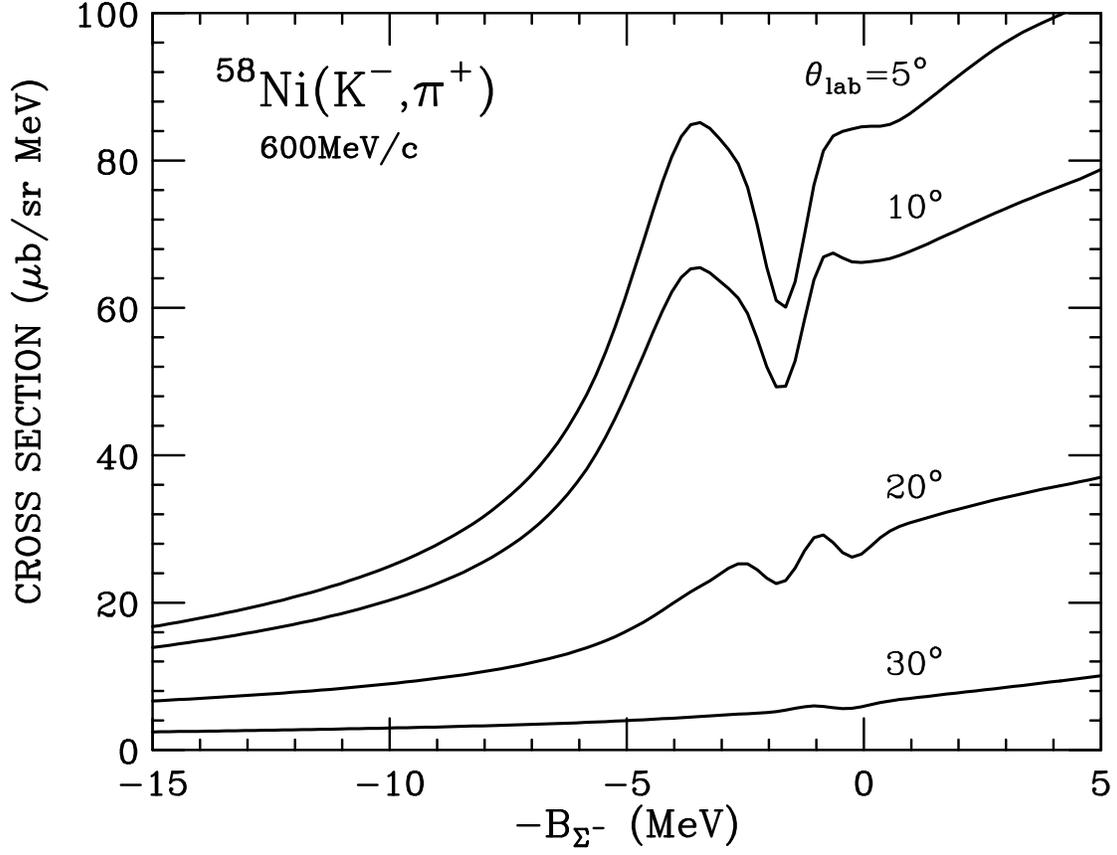}
  \end{center}
  \caption{\label{fig:6}
  Angular dependence of the calculated inclusive 
  spectra of the $^{58}$Ni($K^-$,~$\pi^+$) 
  reaction at $p_K=$ 600 MeV/c near the $\Sigma^-$ threshold, as a 
  function of $-B_{\Sigma^-}$. 
  The curves denote the spectra for $\theta_\text{lab}=$ 5$^\circ$, 
  10$^\circ$, 20$^\circ$ and 30$^\circ$, respectively. 
  The DD potential is used.
  The spectra are folded with a detector resolution of 1.5 MeV FWHM. 
  }
\end{figure}

To see the angular dependence of the spectrum, we calculate 
the inclusive spectra at $p_K=$ 600 MeV/c, 
as a function of $\theta_{\rm lab}$. 
In Fig.~\ref{fig:6}, we show the spectra at 
$\theta_{\rm lab}=$ 5$^\circ$, 
10$^\circ$, 20$^\circ$ and 30$^\circ$, which correspond to 
$q_{\Sigma}\simeq$ 117, 143, 214 and 296 MeV/c, respectively.  
A structure of the peak and dip
below the $\Sigma^-$ threshold in the spectrum disappears 
as increasing $q_\Sigma$. 
The shape of the spectrum at $\theta_{\rm lab}=$ 30$^\circ$ 
in which $q_{\Sigma}\simeq$ 296 MeV/c is near by 
the Fermi momentum of $\sim$270 MeV/c, 
becomes similar to that of the ($\pi^-$,~$K^+$) reaction with 
a high momentum transfer of $\sim$380 MeV/c.
Therefore, the spectrum at $p_K \simeq$ 600 MeV/c (5$^\circ$) 
is expected to be a complement to the analyses of the 
$\Sigma^-$ atomic and ($\pi^-$,~$K^+$) data.

\subsubsection{Comparison between the inclusive spectra for various
 $\Sigma$-nucleus potentials}

\Figuretable{FIG.~7}

\begin{figure}[bth]
  \begin{center}
  \includegraphics[width=0.9\linewidth]{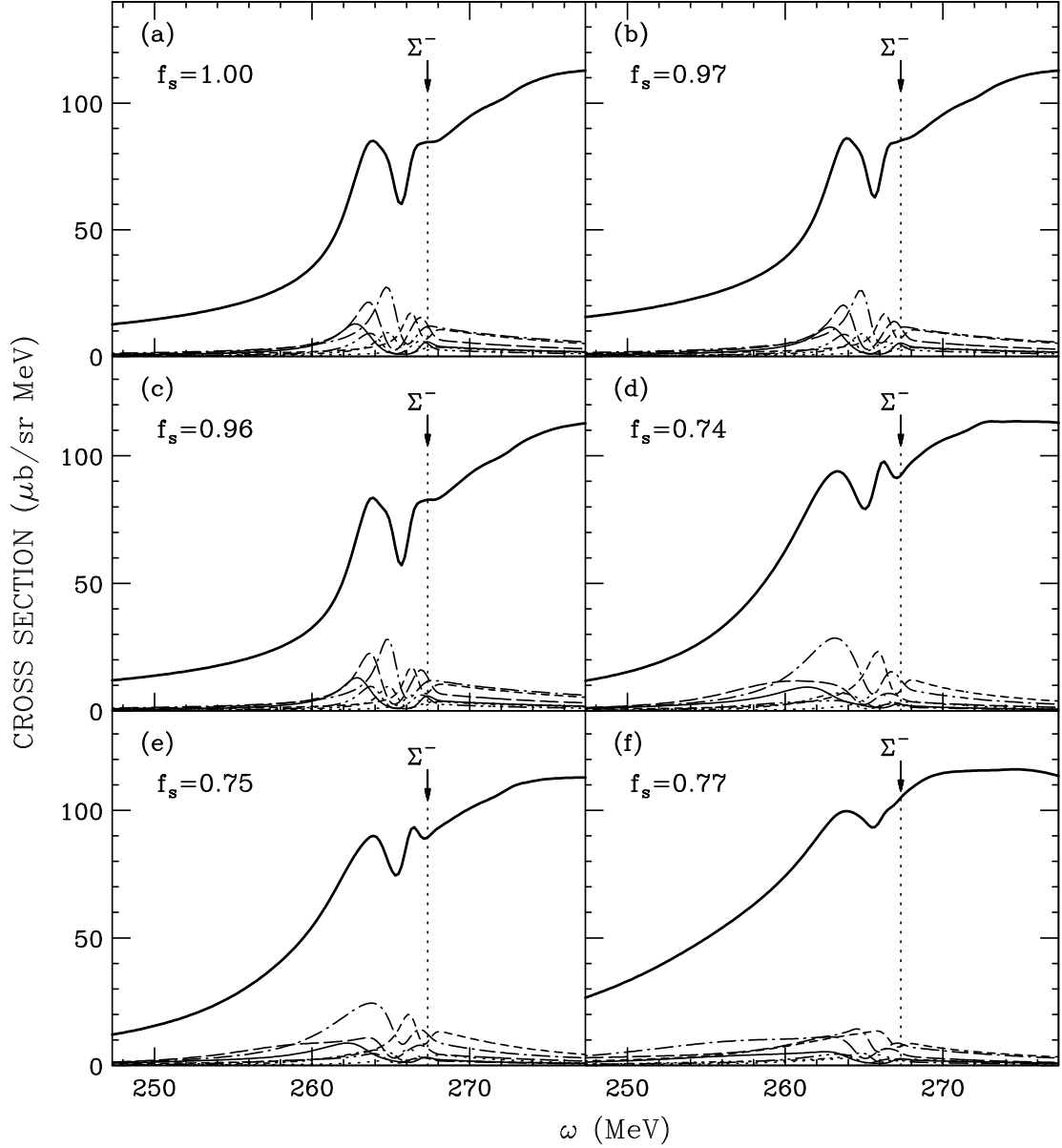}
  \end{center}
  \caption{\label{fig:7}
  Calculated inclusive spectra of the $^{58}$Ni($K^-$,~$\pi^+$) 
  reactions at $p_{K}=$ 600 MeV/c (5$^\circ$) near the $\Sigma^-$ 
  threshold.
  The thick solid curve denotes the spectra with the potentials of 
  (a) DD, (b) RMF, (c) LDA-NF, (d) LDA-S3, (e) WS-sh and (f) $t_{\rm eff}\rho$, 
  where each calculated spectrum is normalized by a factor $f_s$.
  The partial-wave contributions to the inclusive spectra are also 
  drawn (see the Fig.~\ref{fig:5}(c)). 
  The spectra are folded with a detector resolution of 1.5 MeV FWHM.
  The arrows show the $\Sigma^-$+$^{57}$Co threshold energy at 
  $\omega=$ 267.33 MeV. 
  }
\end{figure}

In Fig.~\ref{fig:7}, we show a comparison of the calculated 
spectra of the $^{58}$Ni($K^-$,~$\pi^+$) reaction at 
$p_K= 600$ MeV/c (5$^\circ$), 
using all the types of the potentials mentioned 
in Sect.~\ref{sect:pot}.
For comparison with the shapes of the spectra each other,
we normalize these spectra to the value calculated by DD
at $\omega=$ 277.3 MeV 
(see values of the normalization factors $f_s$ in Fig.~\ref{fig:7}). 
It is clearly shown that the spectra with the potentials having 
a repulsion inside the nucleus such as DD, RMF and LDA-NF, 
are significantly different from those 
with the potentials having some attractions inside 
the nucleus such as LDA-S3, 
WS-sh and $t_\text{eff}\rho$.
The ($K^-$,~$\pi^+$) spectra with a low momentum 
transfer $q_{\Sigma}\simeq$ 110 MeV/c
provides clear evidence to qualitatively identify the  
different type of the $\Sigma$-nucleus potentials.
This fact implies that the near-recoilless ($K^-$,~$\pi^+$) 
spectra confirm the repulsive components 
of the potential inside the nucleus, as a complement to 
$\Sigma^-$ atomic and ($\pi^-$,~$K^+$) data. 
Such a experiment is expected to observe a promising 
spectrum of the nuclear ($K^-$,~$\pi^+$) reaction at J-PARC facilities.

However, we must recognize that 
the shape and magnitude of the spectra with the potentials 
for DD, RMF and LDA-NF are quite similar to each other, 
regardless of a discrepancy between the radial distributions of 
these potentials.
The main reason originates from the appearance of the sizable 
absorption of ${\rm Im}U_\Sigma=$ ($-$20)--($-$30) MeV.
It would be difficult to determine the strength of the repulsion 
at the nuclear center from the inclusive ($K^-$,~$\pi^+$) spectrum.

\subsubsection{Integrated cross sections}

To study the structure of the Coulomb-assisted $\Sigma^-$-nucleus 
bound states, we also calculate the integrated cross sections, 
which can be intuitively understood by the behavior 
of $\Sigma^-$ wave functions, as seen in Eqs.~(\ref{eqn:e54}) 
and (\ref{eqn:e55}). 

\Figuretable{FIG. 8}
\Figuretable{TABLE III}

\begin{figure}[bth]
  \begin{center}
  \includegraphics[width=0.7\linewidth]{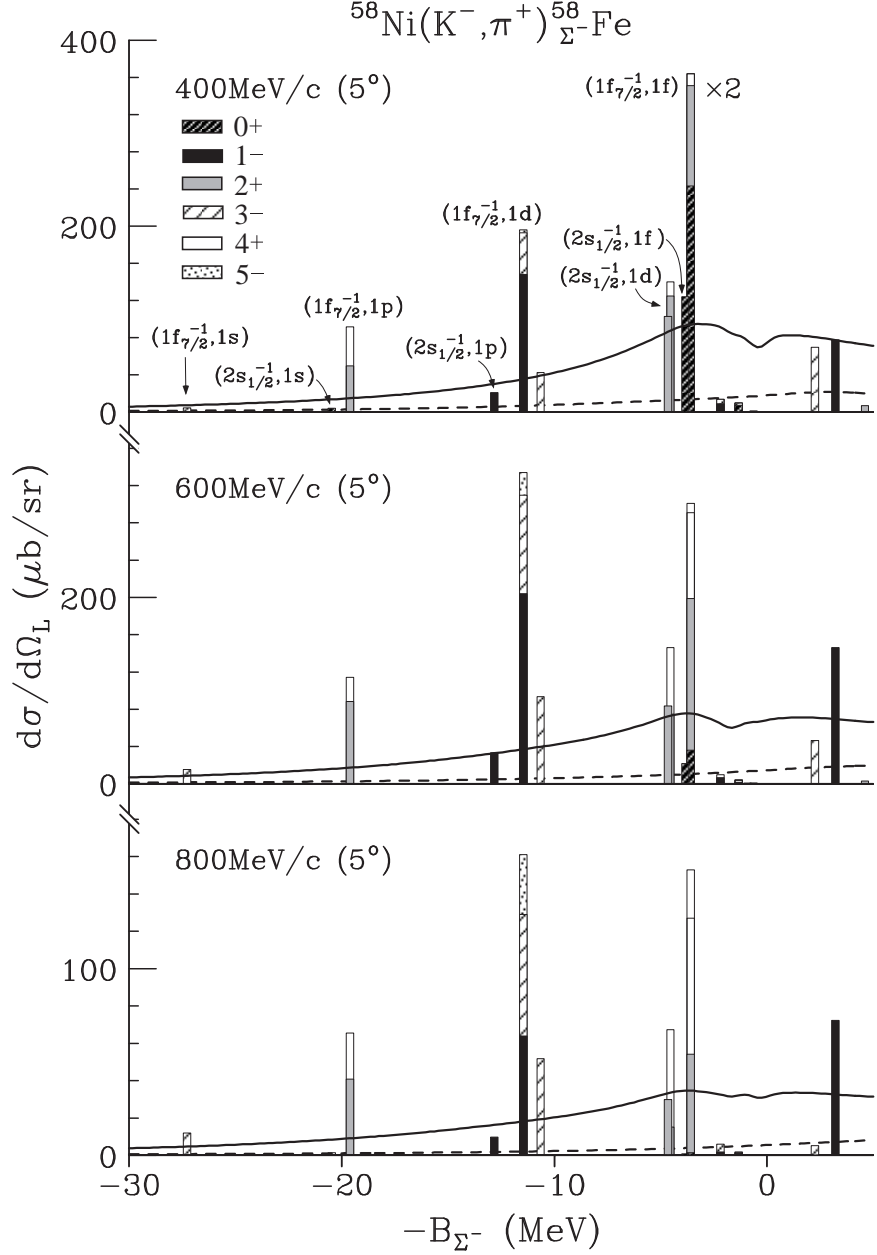}
  \end{center}
  \caption{\label{fig:8}
  Integrated cross sections of the $^{58}$Ni($K^-$,~$\pi^+$) reactions
  at $p_{K}=$ 400, 600 and 800 MeV/c (5$^\circ$) with 
  the $t_\text{eff}\rho$ potential. 
  The values are calculated by the real part of the $\Sigma$-nucleus
  potential including the finite Coulomb one, 
  and its imaginary part is omitted. 
  The solid and dashed curves denote 
  the contributions of the spectra from proton-hole states of 
  $1f_{7/2}^{-1}$ (solid) and $2s_{1/2}^{-1}$ (dash) 
  in the $^{58}$Ni target when the imaginary part of the potential 
  is switched on.
  }
\end{figure}

Firstly, we start to consider the $\Sigma^-$ bound states with 
the $t_\text{eff}\rho$ potential, because 
this situation is analogous to the $\Lambda$-hypernuclear 
production in the ($K^-$,~$\pi^-$) reaction, except their widths. 
In Fig.~\ref{fig:8}, we illustrate the results of 
the integrated cross sections 
of the $\Sigma^-$ bound states with $(j_p^{-1},n\ell_{\Sigma})_{J^{\pi}}$
for $\Sigma^-$--$^{57}$Co in the $^{58}$Ni($K^-$,~$\pi^+$) reactions 
at $p_K=$ 400, 600 and 800 MeV/c (5$^\circ$). 
Here we estimated them with only the real part of the potential 
to clearly see their dependence on $q_\Sigma$ 
when we took into account the $1f_{7/2}^{-1}$ and 
$2s_{1/2}^{-1}$ proton-hole states.

As far as a recoilless kinematics, 
we expect that ``substitutional states'' 
are selectively populated in the $\Sigma^-$ 
bound states, e.g., 
$(1f_{7/2}^{-1},1f_{\Sigma})_{0^+}$ with 
the $\Delta L=$ 0 transition.
This nature approximates to a production at 400 MeV/c 
with $q_\Sigma\simeq$ 66 MeV/c, as seen in Fig.~\ref{fig:8}. 
The components of the $(1f_{7/2}^{-1},1d_{\Sigma})_{3^-,5^-}$ 
states are enhanced as increasing $q_\Sigma$, 
whereas the components of $(1f_{7/2}^{-1},1f_{\Sigma})_{0^+}$ 
are reduced.
We confirm that the partial-wave components in the cross 
section are very sensitive to the value of the momentum transfer.
This behavior is reasonable from the viewpoint of the 
$\Lambda$-hypernuclear production 
\cite{Motoba1988}. 
However, this scenario is changed for the DD potential,
as we will discuss below. 
In Table~\ref{tab:table3}, we list the values of these 
cross sections with $q_\Sigma \simeq$ 117 MeV/c 
at $p_{K}=$ 600 MeV/c (5$^\circ$), 
together with the complex effective numbers of a proton, 
$P_{\rm eff}$. To see the effects of the attraction in the real part 
of the potential, we also show the calculated results 
when we omit the imaginary part of the potential (real only).
We evaluate their argument, ${\rm Re}P_{\rm eff}$ 
and ${\rm Arg}P_{\rm eff}$, in order to understand 
the shape of the inclusive spectrum where 
the imaginary part of the potential is switched on (full).

\Figuretable{FIG. 9}
\Figuretable{TABLE IV}

\begin{figure}[bth]
  \begin{center}
  \includegraphics[width=0.7\linewidth]{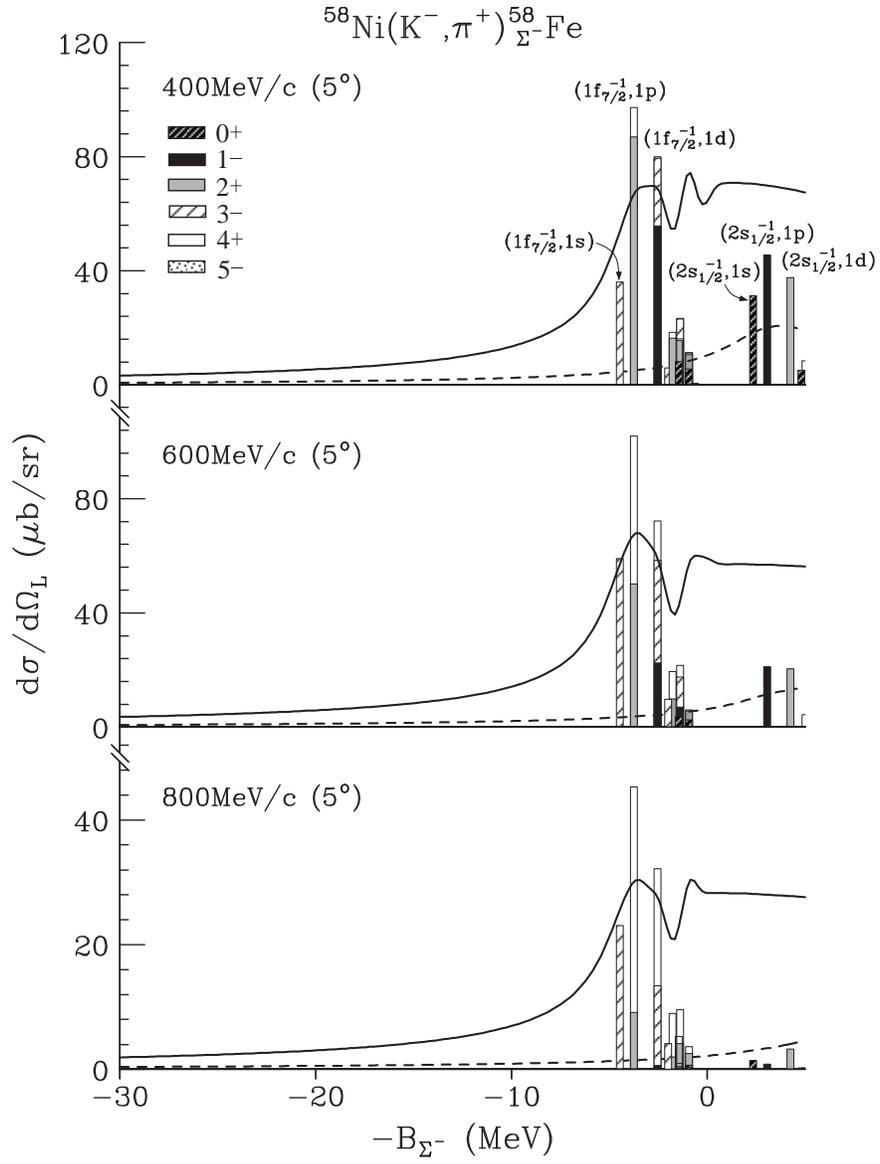}
  \end{center}
  \caption{\label{fig:9}
  Integrated cross sections of the $^{58}$Ni($K^-$,~$\pi^+$) reactions
  at $p_{K}=$ 400, 600 and 800 MeV/c (5$^\circ$) with 
  the DD potential. 
  See the caption in Fig.~\ref{fig:8}. 
  }
\end{figure}

Secondly, we consider the $\Sigma^-$ bound states with the 
DD potential. 
In Fig.~\ref{fig:9}, we illustrate the results of integrated 
cross sections for the $\Sigma^-$ bound states
in $^{58}\text{Ni}$($K^-$,~$\pi^+$) reactions at 
$p_K=$ 400, 600 and 800 MeV/c (5$^\circ$).
We find that the contributions of 
$(1f_{7/2}^{-1},1p_{\Sigma})_{2^+,4^+}$ 
states with $\Delta L=$ 2 are populated dominantly, and 
also $(1f_{7/2}^{-1},1d_{\Sigma})_{1^-,3^-}$ states with 
$\Delta L=$ 1 even if $p_{K}=$ 400 MeV/c. 
This originates predominately from a large difference between 
the wave functions of the $(1d)_\Sigma$ or $(1f)_\Sigma$ state
and the proton-hole $1f_{7/2}^{-1}$ state, 
as seen in Fig.~\ref{fig:4}. 
The high momentum components caused by the repulsion require
a $\Delta L \neq 0$ transition to populate the $\Sigma^-$ bound states.
The distorted waves of $\tilde{j}_L(q,br)$ in Eq.~(\ref{eqn:e16}) 
also moderate the pattern of the population.
In Table~\ref{tab:table4}, we show that ${\rm Re}P_{\rm eff}$ 
is reduced with ${\rm Arg}P_{\rm eff}=$ 45--55$^\circ$ 
in the complex effective number approach \cite{Koike2008}. 
To see the effects of the repulsion in the real part of
the potential, we also list the calculated results without 
the imaginary part of the potential.
Such a peak is shifted downward from the energy position 
at $-B_{\Sigma^-}$
and behaves as an asymmetric shape with a relatively 
narrow width \cite{Morimatsu1994}.

\subsection{$^{28}$Si($K^-$,~$\pi^+$) spectrum}

\Figuretable{FIG. 10}

\begin{figure}[bth]
  \begin{center}
  \includegraphics[width=0.9\linewidth]{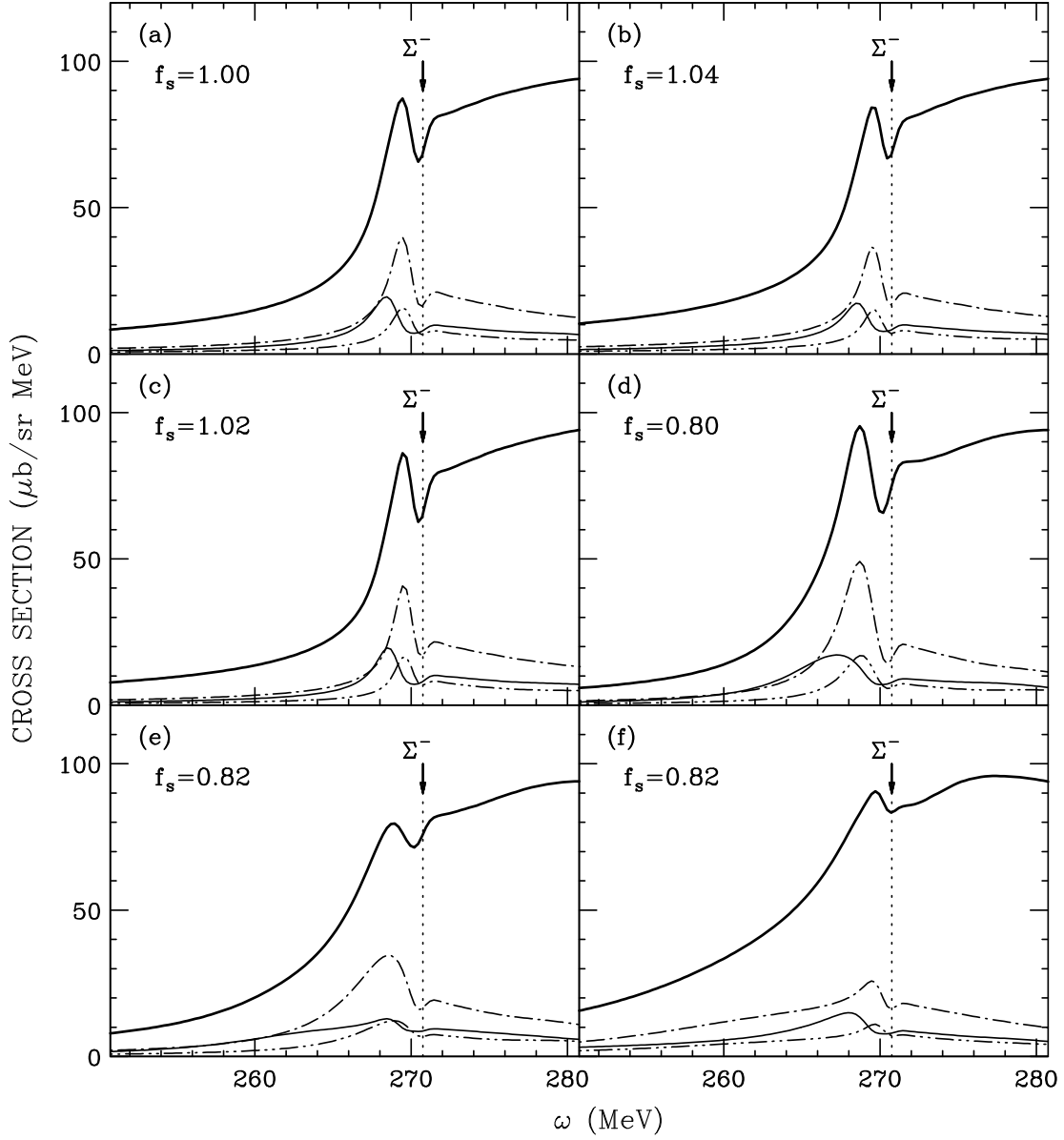}
  \end{center}
  \caption{\label{fig:10}
  Calculated inclusive spectra of the $^{28}$Si($K^-$,~$\pi^+$)
  reactions at $p_{K}=$ 600 MeV/c (5$^\circ$) near the $\Sigma^-$ threshold.
  The thick solid curve denotes the spectra with the potentials of 
  (a) DD, (b) RMF, (c) LDA-NF, (d) LDA-S3, (e) WS-sh and (f) $t_{\rm eff}\rho$, 
  where each calculated spectrum is normalized by a factor $f_s$.
  The solid, dot-dashed and dot-dot-dashed curves denote the 
  partial-wave contributions of $(s_{1/2})_\Sigma$, 
  $(p_{3/2,1/2})_\Sigma$ and $(d_{5/2,3/2})_\Sigma$
  states for $\Sigma^-$--$^{27}$Al, respectively.
  The spectra are folded with a detector resolution of 1.5 MeV FWHM.
  The arrows show the $\Sigma^-$--$^{27}$Al threshold energy at 
  $\omega=$ 270.75 MeV. 
  }
\end{figure}

We attempt to examine a production of the 
$\Sigma^-$ bound states for $\Sigma^-$--$^{27}$Al in the 
$^{28}$Si($K^-$,~$\pi^+$) reaction with a small momentum 
transfer of $q_{\Sigma}\simeq$ 110 MeV/c. 
In Fig.~\ref{fig:10}, we show a comparison of the calculated inclusive 
($K^-$,~$\pi^+$) spectra near the $\Sigma^-$ threshold 
at $600$ MeV/c (5$^\circ$) as a function of the energy 
transfer $\omega$.  
Here we used several types of the potentials that were tested 
by the ($\pi^-$,~$K^+$) reactions in Ref.~\cite{Harada2005}, 
as mentioned in Sect.~\ref{sect:pot}. 
Note that these spectra were renormalized to the value obtained 
by the DD potential at $\omega =$ 280.8 MeV.

We find that a peak of $(1d_{5/2}^{-1},1p_{\Sigma})_{1^-,3^-}$ is
fairly enhanced at $\omega \simeq$ 269 MeV below the $\Sigma^-$ threshold 
in the spectra for the almost types of the potential.
The calculated spectra with the repulsive potentials 
(DD, RMF and LDA-NF)
significantly differ from those with the attractive ones 
(LDA-S3, WS-sh and $t_{\rm eff}\rho$).
Therefore, the discrepancy becomes much more clear 
by comparing the shape behavior of each spectrum in the bound region; 
the absolute values of the former cross sections are 
scarcely smaller than those of the latter ones, 
as seen by the normalization factors $f_s$ in Fig.~\ref{fig:10}.

The $\Sigma^-$ energy levels are pushed up by this repulsion, 
so that the level spacing becomes tight.
Since the binding energy of the $(1p)_\Sigma$ bound 
state is $B_{\Sigma^-}=$ 1.13 MeV for DD in Tables~\ref{tab:table2}, 
which is very different from $B_{\Sigma^-}=$ 9.54 MeV 
for $t_{\rm eff}\rho$, 
the gap energy of the levels between the other bound 
states is not so large.
Consequently, the attraction for 
$\Sigma^-$ with the help of the Coulomb potential
is more needed to clearly identify a $\Sigma^-$ bound state.
We believe that
it is not so difficult to measure a peak structure of 
the Coulomb-assisted $\Sigma^-$-nucleus bound states
in heavier systems such as 
$\Sigma^-$--$^{57}$Co, rather than $\Sigma^-$--$^{27}$Al.

\subsection{$^{208}$Pb($K^-$,~$\pi^+$) spectrum
\label{sect:208Pb}}

To see whether we tell the difference of the repulsive nature 
between DD and LDA-NF or not, 
we consider the spectrum of the Coulomb-assisted 
$\Sigma^-$ bound states in the heavy nuclei.
In Fig.~\ref{fig:2}, we represent the real and imaginary 
parts of the $\Sigma^-$--$^{207}$Tl potentials for DD and LDA-NF, 
together with the finite Coulomb potential.
The Coulomb potential gives a large contribution to the potential 
energy at the nuclear 
center, i.e., $|U_{\rm Coul}| \simeq$ 25 MeV, which is 
comparable to that from the repulsion in LDA-NF. 
As seen in Fig.~\ref{fig:4}, wave functions for the 
$(ns)_\Sigma$, $(np)_\Sigma$ and $(nd)_\Sigma$ states 
are strongly modified by the appearance of the repulsion 
in comparison with the Coulomb wave functions. 
The repulsion in DD at the nuclear center 
(${\rm Re}U_\Sigma \sim 80$ MeV)
is 2--3 times larger than that in RMF or LDA-NF 
(${\rm Re}U_\Sigma \sim 40$ MeV).

\Figuretable{FIG.~11}

\begin{figure}[htb]
  \begin{center}
  \includegraphics[width=0.52\linewidth]{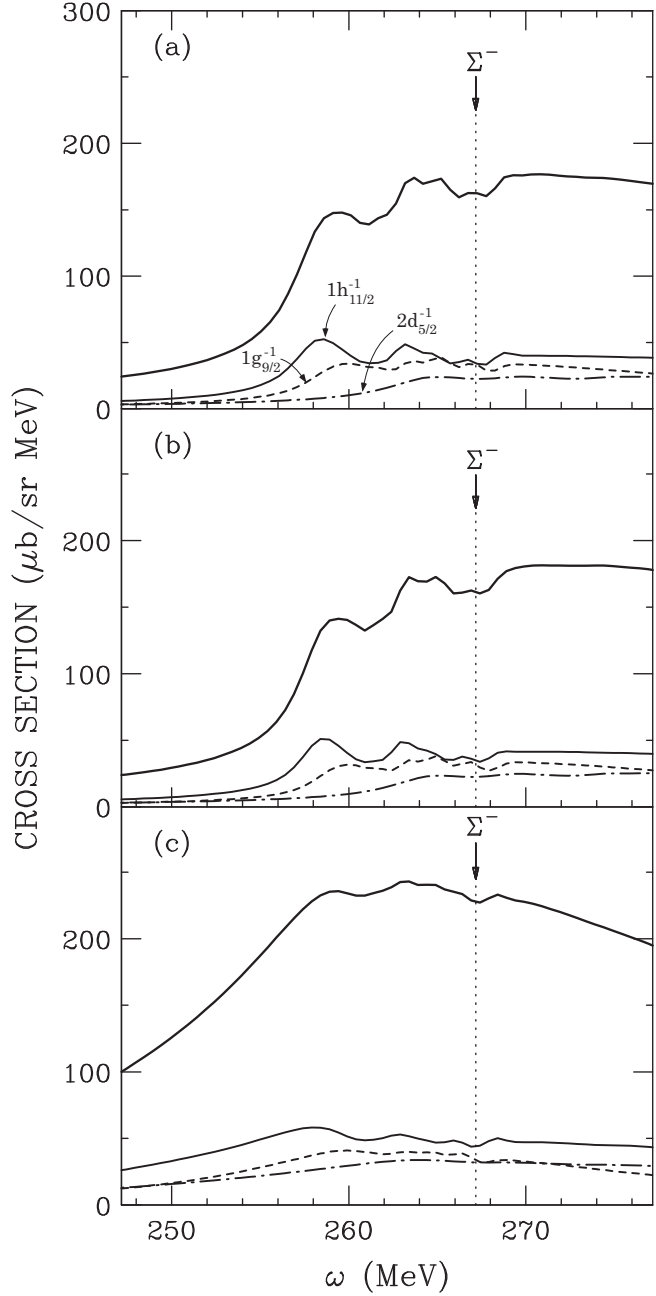}
  \end{center}
  \caption{\label{fig:11}
  Calculated spectra with 16 proton-hole states involving 
  $1h^{-1}_{11/2}$, $2d^{-1}_{5/2,3/2}$ and $1g^{-1}_{9/2,7/2}$ 
  in the $^{208}$Pb($K^-$,~$\pi^+$) reaction 
  at $p_K=$ 600 MeV/c (5$^\circ$) near the $\Sigma^-$ threshold, 
  with the (a) DD, (b) LDA-NF and (c) $t_{\rm eff}\rho$ potentials. 
  The thick solid curves denote the total spectra with each potential. 
  The solid, dash and dot-dashed curves denote the contributions of the 
  $1h^{-1}_{11/2}$, $2d^{-1}_{5/2}$ and $1g^{-1}_{9/2}$ components, 
  respectively.  
  The spectra are folded with a detector resolution of 1.5 MeV FWHM. 
  The arrows show the $\Sigma^-$+$^{207}$Tl threshold energy at 
  $\omega=$ 267.17 MeV. 
  }
\end{figure}

Let us consider the inclusive $^{208}$Pb($K^-$,~$\pi^+$) spectra 
at 600 MeV/c (5$^\circ$). 
Here we use the DD, LDA-NF and $t_{\rm eff}\rho$ potentials.
These types of the potential were also used in calculations of 
$^{209}$Bi($\pi^-$,~$K^+$) reactions \cite{Harada2006a}. 
In Fig.~\ref{fig:11}, we show a comparison between the calculated 
inclusive spectra of $^{208}$Pb($K^-$,~$\pi^+$) reactions, 
taking into account 16 proton-hole states involving $1h_{11/2}^{-1}$, 
$1g_{9/2}^{-1}$ and $2d_{5/2}^{-1}$ proton-hole ones.
We find that the shape of the calculated spectra clearly 
discriminate between the repulsive potentials (DD, LDA-NF)
and the attractive one ($t_{\rm eff}\rho$), but the spectra for DD and 
LDA-NF are very similar to each other. 
The insensitivity of the spectrum to the real part 
of the potential is caused by the following reason.  
The ratio of the expectation values of the $(1s)_\Sigma$ state 
for $\Sigma^{^-}$--$^{207}$Tl, 
\begin{equation}
\left|
{{\langle \text{Re}U_{\Sigma} \rangle} \over 
 {\langle \text{Re}U_{\Sigma} + U_{\rm Coul} \rangle}}
\right|
= 0.025 \quad (0.009)
\end{equation}
in DD (LDA-NF) is suppressed 
rather than 0.131 (0.088)  for $\Sigma^{^-}$--$^{27}$Al. 
Unfortunately, we recognize that the effect of the $\Sigma$-nucleus 
potential is rather masked by the Coulomb potential 
due to the heaver nuclei.

As shown in Fig.~\ref{fig:11}, 
many partial-wave states of the Coulomb-assisted 
$\Sigma^-$-nucleus states can be easily populated 
by the ($K^-$,~$\pi^+$) reaction on $^{208}$Pb.
In contribution from a $1h^{-1}_{11/2}$ proton-hole state,
the first peak at $\omega \simeq 258$ MeV is constructed 
with $(1s)_\Sigma$, $(1p)_\Sigma$ and $(1d)_\Sigma$ states, 
whereas the second peak at $\omega \simeq 263$ MeV 
with $(1h)_\Sigma$, $(2p)_\Sigma$, 
$(2d)_\Sigma$ and $(2f)_\Sigma$ states. 
In $^{208}$Pb, individual levels with broad widths are covering 
each other in ($j_p^{-1}, n\ell_\Sigma$) configurations because the level 
spacing of the populated $\Sigma^-$ states is very tight 
for the Coulomb attraction.
Moreover, the shape and magnitude of these spectra are 
masked by contributions of other proton-hole states \cite{Parkinson1969} 
that can couple with a lot of $\Sigma^-$ states, 
e.g., $3s_{1/2}^{-1}$, $2d_{5/2,3/2}^{-1}$ and $1g_{9/2,7/2}^{-1}$.
This results are contrast to early calculations \cite{Tadokoro1991}
in which only the $1h_{11/2}^{-1}$ proton-hole state was considered.
It should be noticed that 
the contributions of the deep-hole states
are important to examine the spectrum below the $\Sigma^-$ threshold 
quantitatively.
This implies the importance of choosing the appropriate nucleus 
as a target, e.g., $^{58}$Ni, as discussed in Sect.~\ref{sect:58Ni}.

\section{Summary and Conclusion \label{sect:summary}}

We have performed the theoretical study of 
the Coulomb-assisted $\Sigma^-$-nucleus bound states 
to examine properties of the $\Sigma$-nucleus potential
and have attempted to discriminate between 
various types of the $\Sigma$-nucleus potentials
that are consistent with the $\Sigma^-$ atomic X-ray 
and nuclear ($\pi^-$,~$K^+$) data. 
Some potentials we used are repulsive inside 
the nuclear surface and attractive outside the nucleus
with the sizable absorption. 
We have calculated the DWIA inclusive spectra in the 
($K^-$,~$\pi^+$) reactions on $^{28}$Si, $^{58}$Ni 
and $^{208}$Pb targets at $p_K \simeq$ 600 MeV/c 
by using several types of the $\Sigma$-nucleus potential. 
We have predicted the shape and magnitude of the 
inclusive spectra for the Coulomb-assisted $\Sigma^-$-nucleus 
bound states under appropriate kinematics for the forthcoming J-PARC 
experiments. 
The results are summarized as follows:

\begin{itemize}
\item[(i)]  
The nuclear ($K^-$,~$\pi^+$) reaction with the momentum 
transfer of $q_\Sigma\simeq$ 110 MeV/c 
discriminates more clearly between properties of 
the $\Sigma$-nucleus potentials, e.g., repulsive or attractive. 

\item[(ii)] 
The bump structure formed by $(s_{1/2})_\Sigma$, 
$(p_{3/2})_\Sigma$ and $(d_{5/2})_\Sigma$ components 
of $\Sigma^-$--$^{57}$Co is expected to be observed
in the inclusive ($K^-$, $\pi^+$) spectrum on $^{58}$Ni 
at 600 MeV/c (5$^\circ$),
which is shown by the calculation with the repulsive 
$\Sigma$-nucleus potential such as DD.

\item[(iii)]
Details of the radial distributions in the $\Sigma$-nucleus 
potential inside the nucleus and its repulsive component 
at the center are still not determined 
because of the sizable absorption.

\item[(iv)]
For the heavier nuclei, the shape and magnitude 
of the spectrum near the $\Sigma^-$ threshold 
are rather insensitive to the $\Sigma$-nucleus 
potential because the level spacing 
of the populated $\Sigma^-$ states is very tight 
for the Coulomb attraction.

\end{itemize}

\noindent
In conclusion, the near-recoilless ($K^-$,~$\pi^+$) reactions 
on suitable 
nuclear targets such as $^{58}$Ni 
provide a candidate to clearly discriminate between 
qualitative properties of the $\Sigma$-nucleus potentials, 
which can populate the Coulomb-assisted 
$\Sigma^-$-nucleus bound states with a relatively narrow peak
at the forthcoming J-PARC experiments. 
We believe that properties of the $\Sigma$-nucleus potential 
are quantitatively clarified by obtaining valuable information 
on the production of the Coulomb-assisted $\Sigma^-$-nucleus bound 
states in the ($K^-$,~$\pi^+$) reactions, 
as a full complement to the energy-shifts and widths of 
$\Sigma^-$ atomic X-ray data and the $\Sigma^-$  QF spectrum of 
($\pi^-$,~$K^+$) data.

\begin{acknowledgments}

The authors are obliged to Professor A. Gal and 
Dr. J. Mare$\check{\rm s}$ for valuable comments, and are 
also grateful to Dr. T. Koike, Dr. A. Umeya, 
Professor Y. Akaishi and Professor T. Fukuda 
for useful discussion. 
This work is supported by Grant-in-Aid for Scientific Research on 
Priority Areas (No.~17070007).

\end{acknowledgments}


\begin{thebibliography}{99}

\bibitem{Balberg1997}
S. Balberg and A. Gal, Nucl. Phys. {A625} (1997) 435.

\bibitem{Gal1980}
A.~Gal and C.~B. Dover, Phys. Rev. Lett. {44} (1980) 379.


\bibitem{Friedman2007}
E. Friedman and A. Gal, Phys. Rept. {452} (2007) 89, and references therein.


\bibitem{Noumi2002}
H. Noumi, P.~K. Saha, D. Abe, S. Ajimura, K. Aoki, H.~C. Bhang, T. Endo, Y.
  Fujii, T. Fukuda, H.~C. Guo, K. Imai, O. Hashimoto, H. Hotchi, E.~H. Kim,
  J.~H. Kim, T. Kishimoto, A. Krutenkova, K. Maeda, T. Nagae, M. Nakamura, H.
  Outa, M. Sekimoto, T. Saito, A. Sakaguchi, Y. Sato, R. Sawafta, and Y.
  Shimizu, Phys. Rev. Lett. {89} (2002) 072301; 
   {90} (2003) 049902, Erratum.

\bibitem{Saha2004}
P.~K. Saha, H. Noumi, D. Abe, S. Ajimura, K. Aoki, H.~C. Bhang, K. Dobashi, T.
  Endo, Y. Fujii, T. Fukuda, H.~C. Guo, O. Hashimoto, H. Hotchi, K. Imai, E.~H.
  Kim, J.~H. Kim, T. Kishimoto, A. Krutenkova, K. Maeda, T. Nagae, M. Nakamura,
  H. Outa, T. Saito, A. Sakaguchi, Y. Sato, R. Sawafta, and M. Sekimoto, Phys.
  Rev. {C70} (2004) 044613.

\bibitem{kohno2004}
M. Kohno, Y Fujiwara, Y. Watanabe, K. Ogata, and M. Kawai, Prog. Theor. Phys.
  {112} (2004) 895.


\bibitem{Harada2005}
T. Harada and Y. Hirabayashi, Nucl. Phys. {A759} (2005) 143.


\bibitem{Fujiwara2007}
Y. Fujiwara, Y. Suzuki, and C. Nakamoto, Prog. Part. Nucl. Phys. {58} (2007) 439.

\bibitem{Rijken2008}
T.~A. Rijken and Y. Yamamoto, Nucl. Phys. {A804} (2008)  51.

\bibitem{Batty1994}
C.~J. Batty, E. Friedman, and A. Gal, Prog. Theor. Phys. Suppl. 
{117} (1994) 227; Phys. Lett. {B335} (1994) 273.

\bibitem{Harada2006a}
T. Harada and Y. Hirabayashi, Nucl. Phys. {A767} (2006) 206 .

\bibitem{Yamazaki1988}
T. Yamazaki, R.~S. Hayano, O. Morimatsu, and K. Yazaki, Phys. Lett. {B207} (1988)  393.

\bibitem{Dalitz1978}
R.~H. Dalitz and A. Gal, Ann. Phys. (N.Y.) {116} (1978) 167.

\bibitem{Harada1995b}
T. Harada, {\em Proceedings of the 23rd INS International Symposium on Nuclear
  and Particle Physics with Meson Beams in the 1 GeV/c Region, Tokyo, March,
  1995 (Universal Academy Press, Inc., Tokyo, 1995) p. 211} (1995).

\bibitem{Hufner1974}
J. H\"ufner, S.~Y. Lee, and H.~A. Weidenm\"uller, Nucl. Phys. {A234} (1974) 429.

\bibitem{Bouyssy1977}
A. Bouyssy, Nucl. Phys. {A290} (1977)  324.

\bibitem{Auerbach1983}
E.~H. Auerbach, A.~J. Baltz, C.~B. Dover, A. Gal, S.~H. Kahana, L. Ludeking,
  and D.~J. Millener, Ann. Phys. (N.Y.) {148}  (1983)  381.

\bibitem{Dover1980}
C.~B. Dover, L. Ludeking, and G.~E. Walker, Phys. Rev. {C22} (1980) 2073.

\bibitem{Dover1983}
C.~B. Dover and A. Gal, Ann. Phys. (N.Y.) {146} (1983) 309.

\bibitem{Motoba1988}
T. Motoba, H. Band$\bar{\rm o}$, R. W\"unsch, and J. $\check{\rm Z}$ofka, 
Phys. Rev. {C38}  (1988)  1322.


\bibitem{Morimatsu1988}
O. Morimatsu and K. Yazaki, Nucl. Phys. {A483} (1988) 493.

\bibitem{Dover1989}
C.~B. Dover, D.~J. Millener, and A. Gal, Phys. Rep. {184} (1989) 1.


\bibitem{Tadokoro1995}
S. Tadokoro, H. Kobayashi, Y. Akaishi, Phys. Rev. {C51} (1995) 2656.

\bibitem{Koike2008}
T. Koike and T. Harada, Nucl. Phys. {A804}  (2008)  231.

\bibitem{Harada2004}
T. Harada and Y. Hirabayashi, Nucl. Phys. {A744}  (2004)  323.

\bibitem{deVries1987}
H. de~Vries, C.~W. de~Jager, and C. de~Vries, At. Data Nucl. Tables {36}  (1987)  459.

\bibitem{Bohr1969}
A. Bohr and M. Mottelson, {\em Nuclear structure, Vol. 1, p.238.} 
(Benjemin, New York, 1969).

\bibitem{Jacob1966}
G. Jacob and T.A.J. Maris, Rev. Mod. Phys. {38}  (1966) 121.

\bibitem{Dieperink1990}
A.E.L. Dieperink and P.K.A. DeWitt Huberts, Annu. Rev. Nucl. 
Part. Sci. {40} (1990) 239. 


\bibitem{Vautherin1972}
D. Vautherin and D.M. Brink, Phys. Rev. {C5} (1972) 626.


\bibitem{Allardyce1973}
B.~W. Allardyce, C.~J. Batty, D.~J. Baugh, E. Friedman, G. Heymann, M.~E. Cage,
  Pyle~G. J., G.~T.~A. Squier, A.~S. Clough, D.~F. Jackson, S. Murugesu, and V.
  Rajaratnam, Nucl. Phys. {A209}  (1973) 1.

\bibitem{Rosenthal1980}
A.~S. Rosenthal and F. Tabakin, Phys. Rev. {C22}  (1980)  711.

\bibitem{Gopal1977}
G.~P. Gopal, R.~T. Ross, A.~J. van Horn, A.~C. McPherson, E.~F. Clayton, 
T.~C. Bacon, and I. Butterworth, Nucl. Phys. {B119}  (1977)  362.

\bibitem{Dover1979}
C.~B. Dover, A. Gal, G.~E. Walker, and R.~H. Dalitz, Phys. Lett. {89} (1979)  26.


\bibitem{Morimatsu1994}
O. Morimatsu and K. Yazaki, Prog. Part. Nucl. Phys. {33}  (1994)  679.



\bibitem{Berggren1968}
T. Berggren, Nucl. Phys. {A109}  (1968)  265.

\bibitem{Kapur1938}
P.L. Kapur and R. Peierls, Proc. R. Soc. London {A166}  (1938)  277.



\bibitem{Mares1995}
J. Mare\u{s}, E. Friedman, A. Gal, and B.~K. Jennings, Nucl. Phys. {A594} (1995)  311.

\bibitem{Dabrowski2002}
J. D\c{a}browski, J. Ro\`zynek, and G.S. Anagnostatos, Eur. Phys. J. {A14} (2002) 125.


\bibitem{Yamamoto1985}
Y. Yamamoto and H. Band$\bar{\rm o}$, Prog. Theor. Phys. Suppl. {81} (1985) 9;
Y. Yamamoto, T. Motoba, H. Himeno, K. Ikeda, and S. Nagata, Prog. Theor. Phys. Suppl. {117} (1994)  241.


\bibitem{Hayano1988}
R.~S. Hayano, Nucl. Phys. {A478} (1988) 113c.


\bibitem{Gal1981}
A. Gal, G. Toker, and Y. Alexander, Ann. Phys. {137} (1981) 341.


\bibitem{Tadokoro1991}
S. Tadokoro and Y. Akaishi, Phys. Rev. {C42}  (1990) 2591; 
H. Kobayashi, Master's thesis, University of Tokyo, 1995.


\bibitem{Marinov1985}
A. Marinov, W. Oelert, S. Gopal, B. Brinkm\"oller, G. Hlawatsch, 
C. Mayer-B\"oricke, J. Meissburger, D. Paul, M. Rogge, J.~G.~M. R\"omer, 
J.~L. Tain, P. Turek, L. Zemlo, R.~B.~M. Mooy, P.~W.~M. Glaudemans, 
S. Brant, V. Paar, M. Vouk, V. Lopac, 
Nucl. Phys. {A438} (1985) 429.


\bibitem{Parkinson1969}
W.~C. Parkinson, D.~L. Hendrie, H.~H. Duhm, J. Mahoney, J. Saundinos, 
and G.~R. Satchler, Phys. Rev. {178} (1969) 1976.

\end{thebibliography}

\clearpage

\begin{table}[bth]
\caption{
\label{tab:table1}
Energies of single-particle states for a proton in 
$^{28}$Si, $^{58}$Ni and $^{208}$Pb targets, 
which are input in this calculation 
\cite{Jacob1966,Vautherin1972,Dieperink1990}. 
The values in parentheses denote width of a deep-hole state 
for the proton. All energies and widths are in MeV.
}
\begin{ruledtabular}
\begin{tabular}{ccllcllcll}
$(n \ell j)^{-1}_p$  
           && \multicolumn{2}{c}{$^{28}$Si}  
           && \multicolumn{2}{c}{$^{58}$Ni}  
           && \multicolumn{2}{c}{$^{208}$Pb} \\
\hline
$1s_{1/2}$ &&  $-$41 &(10.0) &&  $-$41.9 &(10.0) &&  $-$36.6 &(10.0) \\
$1p_{3/2}$ &&  $-$23 &(6.0)  &&  $-$30.8 &(10.0) &&  $-$33.1 &(10.0)  \\
$1p_{1/2}$ &&  $-$16 &(4.0)  &&  $-$28.9 &(8.0)  &&  $-$32.5 &(10.0)  \\
$1d_{5/2}$ &&$-$11.6 &(0.0)  &&  $-$19.2 &(4.0)  &&  $-$28.4 &(6.0)  \\
$1d_{3/2}$ &&        &       &&  $-$14.5 &(2.0)  &&  $-$27.0 &(6.0)      \\
$2s_{1/2}$ &&        &       &&  $-$14.6 &(2.0)  &&  $-$24.0 &(6.0)      \\
$1f_{7/2}$ &&        &       &&  $-$7.8  &(0.0)  &&  $-$22.9 &(6.0)      \\
$1f_{5/2}$ &&        &       &&        &       &&  $-$20.4   &(4.0)       \\
$2p_{3/2}$ &&        &       &&        &       &&  $-$17.1   &(4.0)      \\
$2p_{1/2}$ &&        &       &&        &       &&  $-$16.0   &(2.0)      \\
$1g_{9/2}$ &&        &       &&        &       &&  $-$15.4   &(2.0)       \\
$1g_{7/2}$ &&        &       &&        &       &&  $-$11.4   &(2.0)      \\
$2d_{5/2}$ &&        &       &&        &       &&  $-$9.7   & (0.0)     \\
$2d_{3/2}$ &&        &       &&        &       &&  $-$9.4   & (0.0)     \\
$3s_{1/2}$ &&        &       &&        &       &&  $-$8.4   & (0.0)      \\
$1h_{11/2}$&&        &       &&        &       &&  $-$8.0   & (0.0)      \\
\end{tabular}                                                           
\end{ruledtabular}                    
\end{table}

\begin{table*}[bth]
\caption{
\label{tab:table2}
Binding energies ($B_{\Sigma^-}$) and widths ($\varGamma_\Sigma$) 
of the $\Sigma^-$-nucleus $(n \ell)_{\Sigma}$ bound states for 
$\Sigma^-$--$^{27}$Al ($^{28}_{\Sigma^-}$Mg),
$\Sigma^-$--$^{57}$Co ($^{58}_{\Sigma^-}$Fe) and
$\Sigma^-$--$^{207}$Tl ($^{207}_{\Sigma^-}$Hg).
Here the $t_{\rm eff}\rho$ and DD 
potentials are used as a $\Sigma$-nucleus potential.
These values are estimated with the $\Sigma$-nucleus 
potential plus the finite Coulomb potential (Full)
and only the finite Coulomb potential (Coulomb). 
$k^\text{(pole)}$ denotes 
a corresponding pole position of the bound state 
in the complex $k$-plane.
}
\begin{ruledtabular}
\begin{tabular}{clccrclrclccr}
       & \multicolumn{4}{l}{$t_{\rm eff}\rho$ (Full)}
       &&\multicolumn{2}{l}{Coulomb} 
       && \multicolumn{4}{l}{DD (Full)}  \\
       \cline{2-5} \cline{7-8}  \cline{10-13}
$(n \ell)_\Sigma$
       & $-B_{\Sigma^-}$ & $\mit\Gamma_{\Sigma}$ 
       & $k^\text{(pole)}$  & rms
       && $-B_{\Sigma^-}$ & rms
       && $-B_{\Sigma^-}$ & $\mit\Gamma_{\Sigma}$  
       & $k^\text{(pole)}$  & rms \\
      & (MeV) & (MeV) & (fm$^{-1}$) &  (fm)  
      && (MeV) & (fm)  
      && (MeV) & (MeV) & (fm$^{-1}$) &  (fm) \\
\hline

$^{28}_{\Sigma^-}$Mg \\
$1s$ &$-$22.0  & 26.9 & $-$0.334+$i$1.186 & 2.1 &&$-$2.96 & 5.2  &&$-$1.71 &1.25 & $-$0.057+$i$0.322 & 9.8    \\
$1p$ &$-$9.54 & 20.4 & $-$0.360+$i$0.831 & 2.9  &&$-$1.25 & 10.4 &&$-$1.13 &0.49 & $-$0.028+$i$0.259 & 12.4   \\
$1d$ &$+$2.26 & 10.6 & $-$0.486+$i$0.322 & 5.9  &&$-$0.57 & 20.4 &&$-$0.58 &0.023 & $-$0.002+$i$0.184 & 20.2  \\
$2s$ &$+$0.26  & 6.12 & $-$0.313+$i$0.287 & 11.9&&$-$0.97 & 15.6 &&$-$0.71 &0.26 & $-$0.019+$i$0.205 & 23.4    \\
$2p$ &$-$0.99 & 0.25 & $-$0.015+$i$0.242 & 14.1 &&$-$0.56 & 25.1 &&$-$0.53 &0.015 & $-$0.013+$i$0.176 & 28.8  \\

\hline
$^{58}_{\Sigma^-}$Fe \\
$1s$ &$-$27.1 & 23.7 & $-$0.273+$i$1.305 & 2.5  &&$-$7.13 & 4.2  &&$-$3.89 &1.89 & $-$0.058+$i$0.488 & 9.1   \\
$1p$ &$-$19.2 & 21.1 & $-$0.286+$i$1.111 & 3.2  &&$-$4.39 & 6.4  &&$-$3.28 &1.43 & $-$0.048+$i$0.447 & 9.8   \\
$1d$ &$-$10.6 & 17.7 & $-$0.311+$i$0.855 & 3.9  &&$-$2.46 & 10.0 &&$-$2.31 &0.63 & $-$0.026+$i$0.373 & 11.7  \\
$1f$ &$-$1.71 & 12.7 & $-$0.382+$i$0.500 & 5.4  &&$-$1.42 & 16.2 &&$-$1.43 &0.080 & $-$0.004+$i$0.293 & 16.2 \\
$2s$ &$-$9.44 & 15.8 & $-$0.294+$i$0.809 & 4.4  &&$-$3.08 & 9.9  &&$-$1.91 &0.49 & $-$0.022+$i$0.340 & 17.9  \\
$2p$ &$-$1.74 & 7.23 & $-$0.261+$i$0.416 & 11.3 &&$-$2.11 & 13.9 &&$-$1.70 &0.42 & $-$0.020+$i$0.320 & 19.3   \\

\hline
$^{208}_{\Sigma^-}$Hg \\
$1s$ &$-$38.4 & 19.1 & $-0.190$+$i1.543$ & 3.5 &&$-$18.80 & 4.2 &&$-$10.5 & 1.7 & $-0.032$+$i0.803$ & 9.8  \\
$1p$ &$-$34.2 & 18.6 & $-0.195$+$i1.459$ & 4.3 &&$-$15.82 & 5.4 &&$-$10.2 & 1.6 & $-0.031$+$i0.788$ & 9.9  \\
$1d$ &$-$29.5 & 17.9 & $-0.201$+$i1.358$ & 4.8 &&$-$12.92 & 6.5 &&$-$ 9.4 & 1.4 & $-0.029$+$i0.758$ & 10.1  \\
$1f$ &$-$24.3 & 16.9 & $-0.209$+$i1.236$ & 5.2 &&$-$10.17 & 7.7 &&$-$ 8.3 & 1.2 & $-0.025$+$i0.714$ & 10.4 \\
$2s$ &$-$28.5 & 17.4 & $-0.199$+$i1.334$ & 4.5 &&$-$13.02 & 6.7 &&$-$ 6.5 & 0.61 & $-0.015$+$i0.633$ & 14.7  \\
$2p$ &$-$22.6 & 16.0 & $-0.205$+$i1.194$ & 5.0 &&$-$10.50 & 8.2 &&$-$ 6.4 & 0.58 & $-0.014$+$i0.623$ & 14.9  \\
$2d$ &$-$16.6 & 14.0 & $-0.208$+$i1.029$ & 5.6 &&$-$8.41  & 10.1&&$-$ 6.0 & 0.53 & $-0.013$+$i0.604$ & 15.3  \\
$3s$ &$-$16.0 & 13.1 & $-0.199$+$i1.008$ & 5.7 &&$-$8.69  & 10.3&&$-$ 4.6 & 0.34 & $-0.010$+$i0.530$ & 20.4  \\
$3p$ &$-$10.3 & 8.30 & $-0.157$+$i0.807$ & 7.3 &&$-$7.17  & 12.4&&$-$ 4.5 & 0.32 & $-0.009$+$i0.523$ & 20.7  \\
\end{tabular}                                            
\end{ruledtabular}
\end{table*}


\begin{table*}[bth]
\caption{
\label{tab:table3}
Integrated lab cross sections of ${\Sigma}^-$-nucleus
bound states ${J^\pi}$ for $\Sigma^-$--${^{57}{\rm Co}}$ 
with the $t_{\rm eff}\rho$ potential, 
by transitions $(n\ell_j)_{p} \to (n\ell_j)_{\Sigma}$ 
in the ($K^-$,~$\pi^+$) reaction on the $^{58}$Ni target
at the incident $K^-$ lab momentum $p_{K}=$ 600 MeV/c and 
$\theta_\text{lab}=5^\circ$. 
The Fermi-averaged cross section of 
$\langle d\sigma/d\Omega \rangle^{K^-p \to \pi^+ \Sigma^-}_{\rm lab}$= 
1.25 mb/sr 
and distortion parameters ${\sigma}_K=$ 32 mb, ${\sigma}_\pi=$ 30 mb and 
$\alpha_K=\alpha_\pi=$ 0 are used in the DWIA.
}
\begin{ruledtabular}
\begin{tabular}{rrrrrlrrrrr}
       && \multicolumn{3}{l}{$t_{\rm eff}\rho$ (Real only)} 
       && \multicolumn{5}{l}{$t_{\rm eff}\rho$ (Full)} \\
       \cline{3-5} \cline{7-11} 
Transition~~~~~~ 
& $J^{\pi}$  & $-B_{\Sigma-}$ & ${\rm Re}P_{\rm eff}$ & ${d\sigma/d\Omega}$ 
             && $-B_{\Sigma^-}$ & $\mit\Gamma_{\Sigma}$ 
& ${\rm Re}P_{\rm eff}$ &  ${\rm Arg}P_{\rm eff}$ & ${d\sigma/d\Omega}$ \\
&            & {(MeV)} & ($\times 10^{-2}$) & ($\mu$b/sr) 
&& {(MeV)}& {(MeV)} 
& ($\times 10^{-2}$) & (deg.) & ($\mu$b/sr) \\ 
\hline
$(1f_{7/2})_p\to (1s_{1/2})_{\Sigma}$
&3$^-$&$-$27.3  &1.218  &15.27  &&$-$27.1 &23.7  &0.928    &$-$33.3   &11.62   \\
$\to (1p_{3/2})_{\Sigma}$                                                                                          
&2$^+$&$-$19.6  &7.254  &88.24  &&$-$19.2 &21.1  &6.129    &$-$25.8   &74.42   \\
&4$^+$&         &0.873  &10.62  &&        &      &0.665    &$-$34.5   &8.07    \\
$\to (1p_{1/2})_{\Sigma}$                                                                                          
&4$^+$&$-$19.6  &1.222  &14.87  &&$-$19.2 &21.1  &0.930    &$-$34.5   &11.30   \\
$\to (1d_{5/2})_{\Sigma}$                                                                                          
&1$^-$&$-$11.5  &17.291 &203.67 &&$-$10.6 &17.7  &16.515   &$-$17.6   &193.84 \\
&3$^-$&         &6.790  &79.98  &&        &      &5.808    &$-$28.5   &68.17   \\
&5$^-$&         &0.599  &7.05   &&        &      &0.502    &$-$31.5   &5.89   \\
$\to (1d_{3/2})_{\Sigma}$                                                                                          
&3$^-$&$-$11.5  &2.263  &26.66  &&$-$10.6 &17.7  &1.936    &$-$28.5   &22.72  \\
&5$^-$&         &1.397  &16.45  &&        &      &1.171    &$-$31.5   &13.74  \\
$\to (1f_{7/2})_{\Sigma}$                                                                                          
&0$^+$&$-$3.58  &3.136  &35.79  &&$-$1.72 &12.7  &4.563    &   30.2   &51.68  \\
&2$^+$&         &12.780 &145.86 &&        &      &15.448   &$-$13.8   &174.99 \\
&4$^+$&         &5.146  &58.73  &&        &      &5.300    &$-$29.1   &60.04  \\                                                                                      
&6$^+$&         &0.235  &2.68   &&        &      &0.282    &$-$23.2   &3.20   \\
$\to (1f_{5/2})_{\Sigma}$                                                                                           
&2$^+$&$-$3.58  &1.534  &17.50  &&$-$1.72 &12.7  &1.854    &$-$13.8   &21.00  \\   
&4$^+$&         &2.859  &32.63  &&        &      &2.945    &$-$29.1   &33.36  \\                                 
&6$^+$&         &0.704  &8.03   &&        &      &0.847    &$-$23.2   &9.60   
\end{tabular}
\end{ruledtabular}
\end{table*}

\begin{table*}[bth]
\caption{
\label{tab:table4}
Integrated lab cross sections of ${\Sigma}^-$-nucleus
bound states ${J^\pi}$ for $\Sigma^-$--${^{57}{\rm Co}}$ 
with the DD potential, 
by transitions $(n\ell_j)_{p} \to (n\ell_j)_{\Sigma}$ 
in the ($K^-$,~$\pi^+$) reaction on the $^{58}$Ni target
at the incident $K^-$ lab momentum $p_{K}=$ 600 MeV/c (5$^\circ$). 
See the caption in Table~\ref{tab:table3}.
}
\begin{ruledtabular}
\begin{tabular}{rrrrrlrrrrr}
       && \multicolumn{3}{l}{DD (Real only)} 
       && \multicolumn{5}{l}{DD (Full)} \\
       \cline{3-5} \cline{7-11} 
Transition~~~~~~ 
& $J^{\pi}$  & $-B_{\Sigma^-}$ & ${\rm Re}P_{\rm eff}$ & ${d\sigma/d\Omega}$ 
             && $-B_{\Sigma^-}$ & $\mit\Gamma_{\Sigma}$ 
& ${\rm Re}P_{\rm eff}$ &  ${\rm Arg}P_{\rm eff}$ & ${d\sigma/d\Omega}$ \\
&            & {(MeV)} & ($\times 10^{-2}$) & ($\mu$b/sr) 
&& {(MeV)}& {(MeV)} 
& ($\times 10^{-2}$) & (deg.) & ($\mu$b/sr) \\ 
\hline
\hline
$(1f_{7/2})_p\to (1s_{1/2})_{\Sigma}$
&3$^-$&$-$4.46  &5.147 &58.95  &&$-$3.89 &1.89  &3.539 & 45.2 &40.45  \\
$\to (1p_{3/2})_{\Sigma}$                                                                                          
&2$^+$&$-$3.74  &4.391 &50.15  &&$-$3.28 &1.43  &2.265 & 55.6 &25.82  \\
&4$^+$&         &1.895 &21.65  &&        &      &1.260 & 45.9 &14.36  \\
$\to (1p_{1/2})_{\Sigma}$                                                                                          
&4$^+$&$-$3.74  &2.653 &30.30  &&$-$3.28 &1.43  &1.764 & 45.9 &20.11  \\
$\to (1d_{5/2})_{\Sigma}$                                                                                          
&1$^-$&$-$2.55  &1.979 &22.49  &&$-$2.31 &0.63  &0.386 & 75.0 &4.38  \\
&3$^-$&         &2.366 &26.89  &&        &      &0.888 & 61.6 &10.09  \\
&5$^-$&         &0.366 &4.16   &&        &      &0.167 & 55.8 &1.90  \\
$\to (1d_{3/2})_{\Sigma}$                                                                                          
&3$^-$&$-$2.55  &0.789 &8.96   &&$-$2.31 &0.63  &0.296 & 61.6 &3.36  \\
&5$^-$&         &0.854 &9.71   &&        &      &0.390 & 55.8 &4.42  \\
$\to (1f_{7/2})_{\Sigma}$                                                                                          
&0$^+$&$-$1.46  &0.008 &0.09   &&$-$1.43 &0.080 &0.010 &$-$6.9 &0.12  \\
&2$^+$&         &0.288 &3.25   &&        &      &0.111 & 56.5 &1.26  \\
&4$^+$&         &0.258 &2.92   &&        &      &0.126 & 47.8 &1.42  \\                                                                                      
&6$^+$&         &0.010 &0.12   &&        &      &0.005 & 49.5 &0.05  \\
$\to (1f_{5/2})_{\Sigma}$                                                                                           
&2$^+$&$-$1.46  &0.035 &0.39   &&$-$1.43 &0.080 &0.013 & 56.5 &0.15  \\   
&4$^+$&         &0.143 &1.62   &&        &      &0.070 & 47.8 &0.79  \\                                 
&6$^+$&         &0.031 &0.35   &&        &      &0.014 & 49.5 &0.16  
\end{tabular}
\end{ruledtabular}
\end{table*}

\clearpage                                                                                               

\end{document}